\documentclass[10pt,notitlepage,superscriptaddress,secnumarabic,amssymb, nobibnotes, aps, pra,nofootinbib,longbibliography]{revtex4-1}

\expandafter\let\csname equation*\endcsname\relax
\expandafter\let\csname endequation*\endcsname\relax
\expandafter\let\csname eqnarray*\endcsname\relax
\expandafter\let\csname endeqnarray*\endcsname\relax
\usepackage{amsmath}
\usepackage{amssymb}
\usepackage{graphicx}

\usepackage[hidelinks]{hyperref}

\newcommand{\RN}[1]{\textup{\uppercase\expandafter{\romannumeral#1}}}

\renewcommand{\rm}{\mathrm}

\begin{document}
\title{Frequency spectrum of an optical resonator in a curved spacetime}

\author{Dennis R\"atzel}
\email{dennis.raetzel@univie.ac.at}
\affiliation{Faculty of Physics, University of Vienna, Boltzmanngasse 5, 1090 Vienna, Austria}

\author{Fabienne Schneiter}
\address{Institut f\"ur Theoretische Physik, Eberhard-Karls-Universit\"at T\"ubingen, 72076 T\"ubingen, Germany}

\author{Daniel Braun}
\address{Institut f\"ur Theoretische Physik, Eberhard-Karls-Universit\"at T\"ubingen, 72076 T\"ubingen, Germany}

\author{Tupac Bravo}
\affiliation{Faculty of Physics, University of Vienna, Boltzmanngasse 5, 1090 Vienna, Austria}

\author{Richard Howl}
\affiliation{Faculty of Physics, University of Vienna, Boltzmanngasse 5, 1090 Vienna, Austria}

\author{Maximilian P.E. Lock}
\affiliation{Imperial College, Department of Physics, SW7 2AZ London, United Kingdom}
\affiliation{Faculty of Physics, University of Vienna, Boltzmanngasse 5, 1090 Vienna, Austria}

\author{Ivette Fuentes}
\affiliation{School of Mathematical Sciences, University of Nottingham, University Park, Nottingham NG7 2RD, UK}
\affiliation{Faculty of Physics, University of Vienna, Boltzmanngasse 5, 1090 Vienna, Austria}

\begin{abstract}
The effect of gravity and proper acceleration on the frequency spectrum of an optical resonator - both rigid or deformable - is considered in the framework of general relativity. The optical resonator is modeled either as a rod of matter connecting two mirrors or as a dielectric rod whose ends function as mirrors. Explicit expressions for the frequency spectrum are derived for the case that it is only perturbed slightly. For a deformable resonator, the perturbation of the frequency spectrum depends on the speed of sound in the rod supporting the mirrors. A connection is found to a relativistic concept of rigidity when the speed of sound approaches the speed of light. In contrast, the corresponding result for the assumption of Born rigidity is recovered when the speed of sound becomes infinite. The results presented in this article can be used as the basis for the description of optical and opto-mechanical systems in a curved spacetime. We apply our results to the examples of a uniformly accelerating resonator and an optical resonator in the gravitational field of a small moving sphere. To exemplify the applicability of our approach beyond the framework of linearized gravity, we consider the fictitious situation of an optical resonator falling into a black hole.\\

%
\end{abstract}


\maketitle

\section{Introduction}

In general relativity (GR), as coordinates have no physical meaning, there is no unique concept for the length of a matter system. Some notion of length can be covariantly defined using geometrical quantities or properties of matter. The ambiguity in the notion of length poses a problem for high accuracy metrological experiments, where gravitational fields or acceleration have a significant role to play. For example, the frequency spectrum of a resonator depends on its dimensions and hence knowledge of the precise values of these dimensions is of utmost importance. Cases in which the effects of gravitational fields and acceleration must be considered include those in which the gravitational field is to be measured, such as in proposals for the measurement of gravitational waves with electromagnetic cavity resonators \cite{Tarabrin:2007en,Reece:1982sc,Pegoraro:1978ont,Pegoraro:1978ele,Grishchuk:1981qua,Grishchuk:1975exc,Gemme:2001ba} or other extended matter systems \cite{Ju:2000det,maggiore2008gravitational,Graham:2012sy,Arvanitaki:2013det,Sabin:2014bua,Goryachev:2014yra,Singh:2016xwa}, tests of GR \cite{Braginskii:1977lab,Howl:2016ryt} or the expansion of the universe \cite{Kopeikin:2014rra,Kopeikin:2013kpa}. Other situations are those in which the metrological system is significantly accelerated \cite{Lock:2016rpe,Lock:2016rmg,Regula:2015cja}. A fundamental limit for the precision of a light cavity resonator as a metrological system can even be imposed by the gravitational field of the light inside the cavity \cite{Braun:2015twa}.

The two most important concepts of length are the proper distance and the radar distance. The proper distance is a geometrical quantity usually associated with the length of a rod that is rigid in the sense of that given by Born \cite{Born:1909the}. The radar distance is the optical length that can be measured by sending light back and forth between two mirrors and taking the time between the two events as a measure of distance. It is this radar length that gives the resonance frequency spectrum of an optical resonator for large enough wave numbers. However, the resonators that are part of the metrological systems described in \cite{Tarabrin:2007en,Reece:1982sc,Pegoraro:1978ont,Pegoraro:1978ele,Grishchuk:1981qua,Grishchuk:1975exc,Gemme:2001ba,Ju:2000det,maggiore2008gravitational,Graham:2012sy,Arvanitaki:2013det,Sabin:2014bua,Goryachev:2014yra,Singh:2016xwa,Braginskii:1977lab,Howl:2016ryt,Kopeikin:2014rra,Kopeikin:2013kpa,Lock:2016rpe,Lock:2016rmg,Regula:2015cja,Braun:2015twa} are confined by solid matter systems, and therefore, the notion of proper length plays also a role. 

In Sec. \ref{sec:rigidrod}, we start our considerations by modeling a 1-dimensional resonator as a set of two end mirrors connected by a rod of matter. If this rod is assumed to be rigid, the resonator is called a rigid optical resonator. In Sec. \ref{sec:fundfreq}, we show that the resonance frequencies of an optical resonator are given by its radar length. The general results derived in Sec. \ref{sec:rigidrod} and Sec. \ref{sec:fundfreq} are applied in the following sections.

Since proper length and radar length are generally different, it turns out that the resonance frequencies of a Born rigid optical resonator change if the resonator is accelerated or is exposed to tidal forces. Furthermore, the frequency of a mode is dependent on the reference time, which, in turn, is dependent on the position of the resonator in spacetime. Taking all this into consideration leads to an expression for the resonance frequencies of a resonator that is dependent on acceleration and curvature. This is presented in Sec. \ref{sec:fundfreqrigid}. 

A realistic rod cannot truly be Born rigid; depending on its stiffness and mass density, it will be affected by the gravitational field and its internal interactions have to obey the laws of relativistic causality. In Sec. \ref{sec:realrod}, we derive expressions for the dependence of the resonance frequencies on the deformation of the rod and show that the change in resonance frequencies depends only on the speed of sound in the material of the rod. Furthermore, we compare the change of the resonance frequencies due to deformations of the rod to the change of the resonance frequencies due to the relativistic effects presented in Sec. \ref{sec:fundfreqrigid}. Additionally, we discuss the notion of a causal rigid resonator which is based on the definition of a causal rigid rod as one composed of a material in which the speed of sound is equivalent to the speed of light. The optical resonator can also be filled with a dielectric, or equivalently, the rod that sets the length of the resonator can be a dielectric material and the mirrors can be its ends. The case of homogeneous isotropic dielectric is discussed in Sec. \ref{sec:dielectric}, and it is shown that the relative frequency shifts are independent of the refractive index of the dielectric material. In Sec. \ref{sec:exampleunif}, we consider the case of a unformly accelerated resonator, in Sec. \ref{sec:examplebh} we consider the case of a resonator that falls into a black hole and in Sec. \ref{sec:examplemass}, we consider the example of an optical resonator in the gravitational field of an oscillating massive sphere. In Sec. \ref{sec:conclusions} we give a summary and conclusions.

In this article, we assume that all effects on the optical resonator can be described as small perturbations. In Sec. \ref{sec:realrod}, we present a certain coordinate system $x^\mathcal{M}$ valid in a region around the world line of the resonator's center of mass in which the spacetime metric takes the form $g_{\mathcal{M}\mathcal{N}}=\eta_{\mathcal{M}\mathcal{N}}+h_{\mathcal{M}\mathcal{N}}$, where $\eta_{\mathcal{M}\mathcal{N}}=\rm{diag}(-1,1,1,1)$ is the Minkowski metric and $h_{\mathcal{M}\mathcal{N}}$ is a perturbation. $h_{\mathcal{M}\mathcal{N}}$ is considered to be small in the sense that $|h_{\mathcal{M}\mathcal{N}}|\ll 1$ for all $\mathcal{M},\mathcal{N}$.

\section{A rigid 1-dimensional resonator in a curved spacetime}
\label{sec:rigidrod}

In GR, the gravitational field is represented by the spacetime metric $g_{\mu\nu}$ on a smooth $4$-dimensional manifold $\mathcal{M}$.  We assume the metric to have signature $(-1,1,1,1)$. Then, for every vector $v^\mu$ at a point $p$ in $\mathcal{M}$, the metric delivers a number $g(v,v)=g_{\mu\nu}v^\mu v^\nu$, which is either positive, zero or negative. These cases are called, respectively, space-like, light-like and time-like. For all space-like vectors $v^\mu$, the square root of the positive number $g(v,v)$ is called the length of this vector. A curve $s(\varsigma)$ parametrized by $\varsigma\in [a,b]$ in the spacetime $\mathcal{M}$ that has tangents $s'(\varsigma):=ds^\mu(\varsigma)/d\varsigma$ that are always space-like is called a space-like curve. The geometrical distance along this curve is the quantity $L_p(s)=\int_a^b d\varsigma \sqrt{g_{\mu\nu} s'^\mu  s'^\nu}$, which is called the proper distance.
To define a frequency we need to know how to measure time. A time measurement in GR is defined only with respect to an observer world line. An observer world line is a curve $\gamma(\varrho)$ whose tangents $\dot\gamma(\varrho):=d\gamma(\varrho)/d\varrho$ are always time-like. The time measured along the observer world line $\gamma(\varrho)$ between the parameter values $\varrho_1$ and $\varrho_2$ is $T_p(\varrho_1,\varrho_2) = \int_{\varrho_1}^{\varrho_2} d\varrho \sqrt{-g_{\mu\nu} \dot \gamma^\mu \dot \gamma^\nu}$. This is the temporal counterpart to the proper distance, and it is called the proper time.
Additionally, at every point of a world line $\gamma(\varrho)$, there is a corresponding set of spatial vectors $v$ called the spatial slice in the tangent vector space at $\gamma(\varrho)$ with respect to $\dot\gamma(\varrho)$, which is defined by the condition $\dot\gamma^\mu(\varrho)v^\nu g_{\mu\nu}(\gamma(\varrho))=0$.

In GR, there exist different notions of rigidity as it turns out to be less than straightforward to formulate this basic concept of Newtonian mechanics in a relativistic way. Early attempts to understand rigidity in the framework of electrodynamics date back to before Einstein's formulation of the special theory of relativity \cite{Abraham:1902pri,Herglotz:1903ele,Schwarzschild:1903ele,Sommerfeld:1904eleI,Sommerfeld:1904eleII}. These approaches turned out to be inconsistent with Lorentz symmetry, which then led to the formulation of a Lorentz invariant differential geometric definition of rigidity in \cite{Born:1909the} by Max Born after special relativity was established. Formulated in a modern way, it is the condition of constant distance between every two infinitesimally separated segments of a rigid body. Here, the measure of distance is the infinitesimal proper distance between the two world lines measured in the spatial slice defined by any of the two world lines. This concept of rigidity is denoted as Born rigidity in literature. A short time after the publication by Born in 1909, it was found by Herglotz \cite{Herglotz:1910sta} and N\"other \cite{Noether:1910kin} that Born rigidity is too restrictive. In particular, they found that, with the exception of the singular case of uniform rotation, the motion of a Born rigid body is completely defined by the trajectory of one of its points. Subsequently, there were attempts to give a less restrictive definition of a rigid body which include the concept of quasi-rigidity in GR, a condition on the multipole-moments of a body \cite{Dixon:1970dynI,Ehlers:1977dyn}, and the model of a rigid body as a body in which the speed of sound is equal to the speed of light \cite{Natario:2014koa}. Here, we will use, as our starting point, a definition of a rigid rod that is Born rigid, and we will undertake a perturbative analysis for small length scales, small accelerations, small velocities and small gravitational fields. In this article, we will show that two types of effects are found; those due to space time properties alone and those due to small deformations of the rod which correspond to small deviations from Born rigidity. Since all effects can be considered to be small, we remain in the linear regime, where the different effects can be assumed to be independent. 

\begin{figure}[h]
\includegraphics[width=6cm,angle=0]{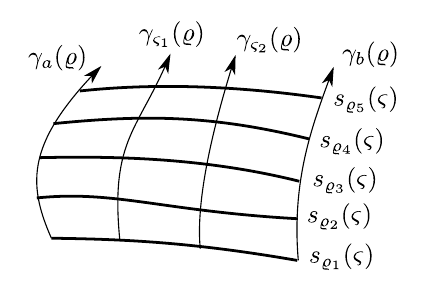}
\caption{\label{fig:rodworldlines} The world lines $\gamma_{\varsigma}(\varrho)$ of the segments of the rod are assumed to form a family of curves which give rise to the rod's world sheet. The curve parameter $\varrho$, which is not necessarily equivalent to their proper time, is the parameter for a family of space-like curves $s_{\varrho}(\varsigma)$ that represent the rod. We assume that the curves $s_{\varrho}(\varsigma)$ are space-like geodesics and cross the world lines of each segment orthogonally.}
\end{figure}
Let us assume that we have a rod of very small diameter in comparison to its length, i.e., it is effectively $1$-dimensional. We assume that the world lines of the segments of the rod form a family of curves $\gamma_\varsigma(\varrho)$ parametrized by $\varsigma$ which we assume to be in the interval $\varsigma\in [a,b]$. The end points of the rod are $\gamma_a(\varrho)$ and $\gamma_b(\varrho)$. The spacetime surface $F(\varrho,\varsigma) = \gamma_\varsigma(\varrho)$ can be called the world sheet of the rod. See Fig. \ref{fig:rodworldlines}. For each curve, the curve parameter $\varrho$ is chosen so that the curves $s_\varrho(\varsigma):=F(\varrho,\varsigma)$ are space-like geodesics in the sense of the auto-parallel condition $\nabla_{s'_\varrho(\varsigma)}s'_\varrho(\varsigma)=0$ with respect to the Levi-Cevita connection $\nabla$ of the metric $g$ given as
$\nabla_\xi\zeta^\alpha=\xi^\beta\partial_\beta\zeta^\alpha+\Gamma^\alpha_{\beta\gamma}\xi^\beta\zeta^\gamma$ for any two vectors $\xi$ and $\zeta$, where
\begin{eqnarray}
	\Gamma^\alpha_{\beta\gamma}
	&=& \frac{1}{2}g^{\alpha\rho}\left(\partial_\beta g_{\gamma\rho}+\partial_\gamma g_{\beta\rho}
		-\partial_\rho g_{\beta\gamma}\right)\,
\end{eqnarray}
are the Christoffel symbols. Note that we do not assume that the world lines of the segments of the rod be geodesics. The segments move under the interior forces of the rod. We also do not assume that $\varrho$ is the proper time of all the segments. Later we will assume that there is a single segment that has $\varrho$ as its proper time. 

For every point of the world sheet $F(\varrho,\varsigma)$ of the rod, we assume that the tangent $s'_\varrho(\varsigma)$ lies in the spatial slice defined by the tangent to the local segment's world line $\dot\gamma_\varsigma(\varrho)$, i.e. $g(\dot\gamma_\varsigma(\varrho),s'_\varrho(\varsigma))=0$. Later, we will find that, due to the condition that the curves $s_\varrho(\varsigma)$ be geodesics, the condition $g(\dot\gamma_\varsigma(\varrho),s'_\varrho(\varsigma))=0$ is fulfilled up to the second order in the proper length of the rod divided by a length scale $l_\rm{var}$, which is associated with local curvature and acceleration. We say that the rod is rigid if the proper distance between every two points on the curve $s_\varrho(\varsigma)$ is independent of the parameter $\varrho$. To further elucidate the meaning of the concept of a rigid rod that we use here, we explain its relation to the concept of a rigid rod that may be familiar from special relativity in Appendix \ref{sec:specialrel}.
\begin{figure}[h]
\includegraphics[width=8cm,angle=0]{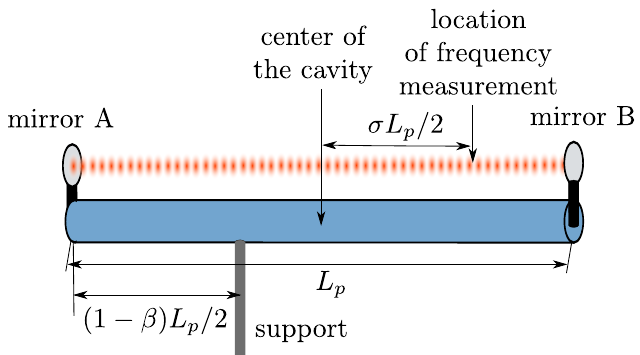}
\caption{\label{fig:rigidcavity} Illustration of our model of an optical resonator consisting of two mirrors that are attached to the ends of a rod. We assume that the resonator is moved along a trajectory $\gamma(\varrho)$ by a support which is attached at a distance $(1-\beta)L_p/2$ from mirror A. Since proper time depends on the position in the gravitational field so does the measured frequency of a resonator mode. We assume the frequency to be measured at a distance $\sigma L_p/2$ from the center of the resonator towards mirror B.}
\end{figure}

There are two possibilities to construct a rigid resonator from the rigid rod defined above. One option is that the rod itself is the resonator: for example, it could be a resonator for electromagnetic waves in different spectral ranges or a resonator for the many different quasiparticles inside and on the surface of a solid matter system such as phonons, plasmons and polaritons, to mention just a few, all of which may resonate between the ends of the rigid rod. The second option is to create a cavity resonator by attaching two mirrors at the endpoints of the rod  such that the light is reflected between the mirrors. In practice, this would be achieved by maximizing the quality factor of the resonator. We denote such resonators as rigid resonators. The second option is the focus of this article, and it is illustrated in Fig. \ref{fig:rigidcavity}. The first option for a homogeneous isotropic dielectric is discussed in Sec. \ref{sec:dielectric}.

A realistic matter system can only be rigid for negligible tidal forces and accelerations. We will discuss our model for a deformable resonator affected by tidal forces and acceleration in Sec. \ref{sec:realrod}. In Sec. \ref{sec:fundfreq}, we will derive an expression for the resonance frequency spectrum of a resonator, rigid or deformable, under the condition that the timescale for light propagation between the mirrors is much smaller than the timescale on which the rigid resonator length is changing.

\section{Resonance frequencies}
\label{sec:fundfreq}

In this section, we will derive an expression for the resonance frequencies of the resonator described above. As we are dealing with an extended object in GR, the obtained resonance frequencies are ambiguous as we will see in the following: first, every mode $k$ existing in the resonator evolves with a certain phase $\psi_k$, this is a covariant quantity. In order to extract a frequency $\omega_k$ from the phase, we require a time $T$ such that we can express the phase as $\psi_k=\omega_k T$. As stated in Sec. \ref{sec:rigidrod}, such a time measurement is defined only with respect to an observer and the time measured by the observer along the curve $\gamma(\varrho)$ is the proper time $T_p(\varrho_1,\varrho_2) = \int_{\varrho_1}^{\varrho_2} d\varrho \sqrt{-g_{\mu\nu} \dot \gamma^\mu \dot \gamma^\nu}$. Through the family of curves associated with the rigid rod, we can define a family of observers along the curves $\gamma_\varsigma(\varrho)=s_\varrho(\varsigma)$. We see that every point in the resonator corresponds to a different observer and, therefore, we cannot give a proper time to the whole resonator, therefore the frequencies of the modes must depend on the point in the resonator where they are observed. 

First, we will consider the case of an optical resonator, discussing other cases at the end of the section. The resonance frequencies can be obtained from the evolution of the phase $\psi_k$ of a resonator mode. This can be found by explicitly solving Maxwell's equations in the curved spacetime under consideration. However, we can achieve the same result much faster by implementing the short wavelength expansion or geometric optical limit. The purpose of the following calculation is to prove the expression in Eq. \ref{eq:fundfreq}, which gives the resonance frequencies in terms of the radar distance between the two ends of the resonator. Some readers may want to jump to Eq. \ref{eq:fundfreq} directly. 

In the short wavelength expansion, the electromagnetic field strength tensor for a freely propagating, monochromatic light wave is given as \cite{Perlick:2000ray}
\begin{equation}\label{eq:Amueikonal}
	F_{\mu\nu}(x)=\mathrm{Re}\left(e^{i \frac{\alpha}{\lambda} S(x)}\sum_{n=0}^{\infty} \phi_{n,\mu\nu}(x) \left(\frac{\lambda}{\alpha}\right)^n\right) \,,
\end{equation}
where the complex valued second rank tensors $\phi_{n,\mu\nu}(x)$ give the slowly varying amplitudes, $\lambda$ is the wavelength, $\alpha$ is the length scale of the slow changes of the properties of the light field and the real function $S(x)$ is the eikonal function which describes the rapidly varying phase. In particular, $\alpha$ is the smallest of the length scales given by the waist of the resonator mode, the acceleration of the cavity and the spacetime curvature. 
This statement will get its full meaning in Sec. \ref{sec:realrod}, where the effects of the motion of the resonator and the spacetime curvature on the proper length of the resonator are considered explicitly by using a particular set of coordinates called the proper detector frame. We assume that $\lambda\ll \alpha$ and $\lambda < L_p$. We will only consider linear polarization in the following. We find that the results for the change of the frequency spectrum do not depend on the polarization. Therefore, the results also apply to circular and elliptic polarized fields as those can be obtained as superpositions of linearly polarized fields.

The raised gradient of the eikonal function $\hat \xi^\mu(x):=g^{\mu\nu}\partial_\nu S(x)$ is the normal vector field to the wave fronts defined by $S(x)$. Applying the Maxwell equations to the eikonal expansion in Eq. (\ref{eq:Amueikonal}), we find in leading order that $\hat \xi^\mu(x)$ must be a light-like vector field, i.e. $\hat \xi^\mu(x)\hat \xi^\nu(x)g_{\mu\nu}(x)=0$ \cite{Straumann:2012gen} \footnote{For any matter field in the eikonal approximation, the gradient of the eikonal function has to fulfill the characteristic equations which derive from the highest derivative part of the matter field equations. In the case of Maxwells electrodynamics, the characteristic equations are simply given by the light cone condition. For more details about this analysis see \cite{Rivera2012:thesis,Perlick:2000ray,Audretsch:1991est}.}. Additionally, the light like condition implies that the integral curves of the tangents $\hat \xi^\mu(x)$ are light-like geodesics. In other words, there exist curves $\xi(\varsigma)$ that have the tangents $\hat \xi^\mu(\xi(\varsigma))$: the light rays of geometric optics. Furthermore, the light like property implies $\hat \xi^\mu(x)\partial_\mu S(x)=0$, which means that the phase $\frac{\alpha}{\lambda}S(x)$ is constant along the light rays. We will use these properties of the eikonal function and its gradient to derive the frequency spectrum of the optical resonator in the following.

Inside a resonator, we create standing waves. Hence, we must assume that, for the resonator, there are stationary solutions of Maxwell's equations that fulfill the boundary conditions at the mirrors.  This assumption is valid if we assume that coordinates exist in a small region containing the resonator such that the positions of the mirrors and the metric change only very slightly in the time span that light needs to propagate between the mirrors.
Assuming that linearly polarized standing cavity mode solutions exist, we consider the superposition of two counter-propagating linearly polarized light waves $F_{\mu\nu}^{\mathrm{res}}(x)=F_{\mu\nu}^{r}(x) + F_{\mu\nu}^{l}(x)$, where $F_{\mu\nu}^{r}(x)$ and $F_{\mu\nu}^{l}(x)$ are as in Eq. (\ref{eq:Amueikonal}) with the eikonal functions $S^r(x)$ and $S^l(x)$, respectively. $F_{\mu\nu}^{l}(x)$ represents the wave propagating to the left (negative direction) and $F_{\mu\nu}^{r}(x)$ represents the wave propagating to the right (positive direction). We obtain 
\begin{equation}\label{eq:Amueikonalstanding}
	F_{\mu\nu}^{\mathrm{res}}(x)=\mathrm{Re}\left( e^{i \frac{\alpha}{\lambda} S^r(x)}\sum_{n=0}^{\infty} \phi^r_{n,\mu\nu}(x) \left(\frac{\lambda}{\alpha}\right)^n + e^{i \frac{\alpha}{\lambda} S^l(x)}\sum_{n=0}^{\infty} \phi^l_{n,\mu\nu}(x) \left(\frac{\lambda}{\alpha}\right)^n\right) \,.
\end{equation}
\begin{figure}[h]
\includegraphics[width=8cm,angle=0]{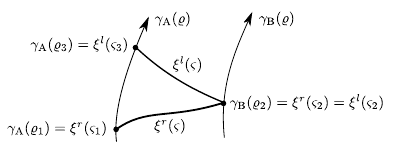}
\caption{\label{fig:pathlight}
The resonance frequencies of a resonator can be derived in the geometric optical limit by considering light bouncing back and forth between the two mirrors of the optical resonator.}
\end{figure}

We defined a rigid cavity by assuming that there are two mirrors attached to the ends of a rigid rod. We consider the gravitational attraction of the two mirrors, all atoms in the rigid rod and the light itself to be negligible. We assume in the following that the mirrors are so close to the ends and so tightly attached that we can identify their world lines with those of the end points of the rod, i.e. $\gamma_\rm{A}(\varrho)=\gamma_a(\varrho)$ and $\gamma_\rm{B}(\varrho)=\gamma_\rm{b}(\varrho)$. Starting at $\varrho=\varrho_1$ with the mirror at $\gamma_\rm{A}(\varrho_1)$, we can define a curve $\xi^r(\varsigma)$ with $\varsigma\in [\varsigma_1,\varsigma_2]$ such that $\xi^r(\varsigma_1)=\gamma_\rm{A}(\varrho_1)$ and $\xi^r(\varsigma_2)=\gamma_\rm{B}(\varrho_2)$ for some $\varrho_2$ and $d\xi^{r,\mu}(\varsigma)/d\varsigma=\hat \xi^{r,\mu}(\xi(\varsigma))=g^{\mu\nu}\partial_\nu S^r(\xi(\varsigma))$ (see Fig. \ref{fig:pathlight} for an illustration). Since all tangents of $\xi^r(\varsigma)$ are light like, this is a light-like curve and can be interpreted as the path of a massless point particle, a single photon, from mirror A to mirror B. At mirror B, the photon is reflected and the tangent of its path becomes $g^{\mu\nu}\partial_\nu S^l(\gamma_\rm{B}(\varrho_2))$. We can define a curve $\xi^l(\varsigma)$ with $\varsigma\in [\varsigma_2,\varsigma_3]$ such that $\xi^l(\varsigma_2)=\gamma_\rm{B}(\varrho_2)$ and $\xi^l(\varsigma_3)=\gamma_\rm{B}(\varrho_3)$ for some $\varrho_3$ and $d\xi^{l,\mu}(\varsigma)/d\varsigma=\hat \xi^{l,\mu}(\xi(\varsigma))=g^{\mu\nu}\partial_\nu S^l(\xi(\varsigma))$. This is the light-like curve representing the path of the photon back to the mirror A. At mirror A, the photon is again reflected and the tangent becomes $g^{\mu\nu}\partial_\nu S^r(\gamma_\rm{A}(\varrho_3))$. 

Then, a condition can be formulated that is necessary to fulfill the boundary conditions at each of the mirrors: the phases of the left propagating and the right propagating parts of $F_{\mu\nu}^{\mathrm{res}}(\gamma_\rm{A}(\varrho))$ and $F_{\mu\nu}^{\mathrm{res}}(\gamma_\rm{B}(\varrho))$ have to match by a multiple of $2\pi$. In Appendix \ref{sec:boundary}, the derivation of this condition is given.
Since the phase is constant along the geodesics $\xi^r$ and $\xi^l$, we find that the change of the eikonal function at the position of the mirror must have been $\frac{\alpha}{\lambda}\delta S_\rm{A}=\frac{\alpha}{\lambda}(S(\gamma_\rm{A}(\varrho_3))-S(\gamma_\rm{A}(\varrho_1)))=2\pi m$ where $m\in \mathbb{Z}$. An observer at mirror A can measure this phase and associate it with a frequency and a change in proper time as $\frac{\alpha}{\lambda}\delta S_\rm{A} = \omega_\rm{A} T_p(\varrho_1,\varrho_3)$. The proper time difference $T_p(\varrho_1,\varrho_3)$ is proportional to the radar length $R_\rm{A}=c T_p(\varrho_1,\varrho_3)/2$ of the resonator measured at $\varrho_0=(\varrho_3+\varrho_1)/2$ by an observer traveling with mirror A. Therefore, we find that the frequencies of the modes of the resonator measured by an observer along the world-line of mirror A are given as
\begin{equation}\label{eq:fundfreqa}
	\omega_{\rm{A},n} = \frac{c n\pi}{R_\rm{A}}\,,
\end{equation}
where we assume $n>0$, i.e. we consider only positive frequencies.
A similar analysis can be made for mirror B, which leads to $\omega_{\rm{B},n} = \frac{c n\pi}{R_\rm{B}}$. Accordingly, for any other observer inside the cavity, we obtain
\begin{equation}\label{eq:fundfreq}
\omega_{\gamma,n}=\frac{c n\pi}{R_{\gamma}}\,,
\end{equation}
where $R_{\gamma}$ is obtained by following a light like geodesic from the observer to one of the mirrors, after reflection, to the second mirror and, after the second reflection, back to the observer. It is clear that this is an approximate value; the notion of frequency means the rate of repetition of a signal. For this notion to make sense, it has to be constant at least for a few repetition cycles. Hence, the observer measuring the frequency has to move slowly in comparison to the time that a light pulse needs to propagate between the mirrors $R_{\gamma}/c$. 

There is another way to understand Eq. (\ref{eq:fundfreq}): electrodynamics in a Lorentzian spacetime can be interpreted as electrodynamics in a non-dispersive, bi-anisotropic, impedance matched medium using the Plebanski constitutive equations \cite{Plebanski1960}
\begin{eqnarray}\label{eq:plebanski}
	D^i &=& \varepsilon_0\varepsilon^{ij}E_j + \frac{1}{c}\epsilon^{ijk}w_jH_k\,,\\
	B^i &=& \mu_0\mu^{ij}H_j -\frac{1}{c}\epsilon^{ijk}w_jE_k\,,
\end{eqnarray}
where we define the spatial co-vector as $w_i:=g_{i0}/g^{00}$ and the permittivity and permeability matrices \\
$\varepsilon^{ij}=\mu^{ij}:=-\sqrt{|\det{g}|}\,g^{ij}/g_{00}$. Maxwell's equations in the curved spacetime $g_{\mu\nu}$ take the form of Maxwell's equations in this effective dielectric medium in flat spacetime. Note that the spatial co-vector $w_j$, which mixes the electric and magnetic field components, is defined by the space-time mixing components of the metric. If the metric is orthogonal in the chosen set of coordinates, $w_j$ vanishes and we are left with a normal anisotropic medium.

Let us assume that the coordinate system was chosen such that the coordinate time $t$ coincides with the proper time at mirror A and that $z$ is the coordinate along the light ray. In this case, we find that the radar length of the resonator measured by an observer at mirror A can be written as
\begin{eqnarray}
	 \nonumber R_\rm{A} &=& \frac{c}{2}(t_2-t_1) = \frac{c}{2}\int_{t_1}^{t_2} dt' = c\int_{z_a}^{z_b} \left(\frac{dz}{dt}\right)^{-1} dz \\
	\label{eq:opticalpath}
	&=& \int_{z_a}^{z_b} \frac{c}{v_\rm{ph}} dz = \int_{z_a}^{z_b} n_z dz\,,
\end{eqnarray}
where $v_\rm{ph}=dz/dt$ is the coordinate dependent phase velocity of the light and $n_z=c/v_\rm{ph}$ can be understood as an effective index of refraction. Eq. (\ref{eq:opticalpath}) shows that the radar length can be understood as the optical path length measured by a ray sent from mirror A to mirror B. Hence, Eq. (\ref{eq:fundfreq}) is the condition that the frequencies measured at mirror A must be multiples of the speed of light divided by the optical path length.

At the end of this section, we would like to discuss the effect of higher order terms in the eikonal expansion. We derived the frequency spectrum (\ref{eq:fundfreqa}) and (\ref{eq:fundfreq}) from a necessary condition for the existence of linearly polarized standing wave solutions of the electromagnetic field in the resonator. This is the condition at the leading order in the eikonal expansion. Terms in the eikonal expansion of higher order may be complex functions in general, this can lead to additional phase shifts at the boundaries which, in turn, can lead to frequency shifts. Such additional frequency shifts can be either considered as systematical errors that limit the predictive power of our approach or have to be evaluated independently to be subtracted from the result of the measurement. One particular source of additional frequency shifts is rotation of the resonator about an axis orthogonal to its optical axis. For earthbound experiments, such rotation will be induced by the rotation of the Earth, for example, which can be measured independently and taken into account explicitly. The effect of rotation may be calculated by taking higher orders of the eikonal expansion into account or using other methods of electrodynamics such as the paraxial wave equation. Here we assume that the optical resonator is non-rotating and we restrict our considerations to the expression for the frequency spectrum given in Eq. (\ref{eq:fundfreq}). In the next section, we will look at its application.


\section{Born rigid optical resonators}
\label{sec:fundfreqrigid}

In this section, we will derive the resonance frequencies of a Born rigid resonator in terms of its constant proper length. For this purpose, we choose to work in a particular coordinate system which we will introduce in the following.

\begin{figure}[h]
\includegraphics[width=6cm,angle=0]{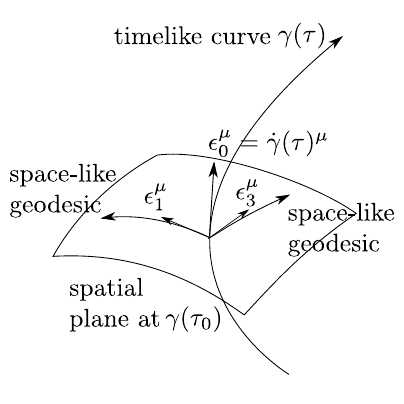}
\caption{\label{fig:properdetectorframe} The proper detector frame can be defined along any time-like curve $\gamma$. The time coordinate is the proper time $\tau$ measured along the curve. The spatial coordinates at a proper time $\tau_0$ are constructed from the proper distances along space-like geodesics that originate at $\gamma(\tau_0)$. The point with coordinates $(c\tau_0,x,y,z)$ is found by following the spatial geodesic with tangent $x^a\epsilon_a^\mu$ a proper distance $(x^2+y^2+z^2)^{1/2}$ from $\gamma(\tau_0)$.}
\end{figure}

Along the world line of an observer $\gamma(\tau)$, an orthonormal, co-rotating tetrad $\epsilon_\mathcal{M}^\mu(\tau)$ ($\mathcal{M}\in\{0,1,2,3\}$, all calligraphic capital letters will run from $0$ to $3$ in the following) can be defined where $\epsilon_0^\mu=\dot \gamma^\mu(\tau)$ is the tangent to the world-line of the observer, $\epsilon_J^\mu(\tau)$ ($J\in\{1,2,3\}$, all capital non-calligraphic letters will run from $1$ to $3$ in the following) are space-like, $\epsilon_\mathcal{M}^\mu(\tau)\epsilon_\mathcal{N}^\nu(\tau)g_{\mu\nu}(\gamma(\tau))=\eta_\mathcal{MN}$ and $\eta_\mathcal{MN}=\rm{diag}(-1,1,1,1)$. There also exists a corresponding co-tetrad $\varepsilon_\mu^\mathcal{M}$ with $\varepsilon_\mu^\mathcal{M}\epsilon_\mathcal{N}^\mu=\delta^{\mathcal{M}}_\mathcal{N}$. The proper distance along the space-like geodesics extending from $\gamma(\tau)$ in the spatial directions generated from $\epsilon_J^\mu(\tau)$ and the proper time $\tau$ along the world line of the observer generate a coordinate system that is associated with the observer (see Fig. \ref{fig:properdetectorframe}). This coordinate system only exists in the vicinity of the observer's world line, as it can only here be ensured that the spatial hyper-planes generated by $\epsilon_J^\mu(\tau)$ at different $\tau$ do not intersect. In these coordinates, the spacetime metric seen by a non-rotating observer can be given simply in terms of: the Riemann curvature tensor along $\gamma(\tau)$ given as $R_\mathcal{MNKL}(\tau)=\epsilon_\mathcal{M}^\alpha(\tau)\epsilon_\mathcal{N}^\beta(\tau)\epsilon_\mathcal{K}^\gamma(\tau)\epsilon_\mathcal{L}^\delta(\tau) g_{\alpha\sigma}(\gamma(\tau))R_{\alpha\beta\gamma\delta}(\gamma(\tau))$ where
\begin{eqnarray}
	{R^\alpha}_{\beta\gamma\delta}
	&=& \partial_\gamma \Gamma^\alpha_{\beta\delta} - \partial_\delta \Gamma^\alpha_{\beta\gamma}
		+\Gamma^\alpha_{\gamma\rho} \Gamma^\rho_{\beta\delta} -\Gamma^\alpha_{\delta\rho} \Gamma^\rho_{\beta\gamma}	\,;
\end{eqnarray}
and the non-gravitational acceleration with respect to a local freely falling frame, represented by the spatial vector $\mathbf{a}^J:=\varepsilon_\mu^J a^{\mu}$, where $a^{\mu}=(\nabla_{\dot\gamma}\dot\gamma)^\mu$.

This coordinate system is called {\it Fermi normal coordinates} for a freely falling, non-rotational observer ($\mathbf{a}=0$) \cite{manasse1963fermi} or {\it the proper detector frame} if proper acceleration occurs \cite{Ni1978proper,maggiore2008gravitational}. The proper detector frame of a non-rotating observer is accurate for proper distances \cite{Ni1978proper}
\begin{equation}
	|\mathbf{x}|\ll l_\mathrm{var}=\mathrm{min}\left\{\frac{c^2}{|\mathbf{a}^J|},\frac{1}{|{R^\mathcal{M}}_{\mathcal{N}\mathcal{P}\mathcal{Q}}|^{1/2}},\frac{|{R^\mathcal{M}}_{\mathcal{N}\mathcal{P}\mathcal{Q}}|}{|{R^\mathcal{M}}_{\mathcal{N}\mathcal{P}\mathcal{Q},\mathcal{R}}|}\right\}\,.
\end{equation}
In the following, we will assume that the length of the resonator $L_p$ is small in comparison to the scale $l_\mathrm{var}$. We consider $\gamma(\tau)$ to be the world-line of the point at which the rod that holds the resonator is supported. We assume that this point is somewhere inside the resonator. If it is not attached to any device, we assume that the center of acceleration is the rod's center of mass. We also assume that the resonator is not rotating in the frame of the observer. We orient the spatial geodesic representing the rigid rod along the $z$-direction at $\gamma(\tau)$, i.e. $s'_\tau(\varsigma)=(0,0,0,1)$. By construction of the proper detector frame, the geodesics $s_\tau(\varsigma)$ run along the $z$-coordinate. Then, we assume that 
\begin{equation}\label{eq:curvdom}
	l_\mathrm{var}=\mathrm{min}\left\{\frac{1}{|{R^0}_{z0z}|^{1/2}}\right\}\ll \mathrm{min}\left\{\frac{c^2}{|a^J|}\right\} \,.
\end{equation}
In particular, Eq. (\ref{eq:curvdom}) allows us to neglect all higher order contributions of acceleration. With this assumption, we can consider the metric in the proper detector frame as a linearly perturbed flat spacetime metric. We define the metric perturbation $h^{P}_{\mathcal{M}\mathcal{N}}:= g^{P}_{\mathcal{M}\mathcal{N}}-\eta_{\mathcal{M}\mathcal{N}}$. For example, in the gravitational field of the earth, the inverse of the square root of the spatial curvature in the direction away from the center of the earth is of the order of $10^{11}\,\rm{m}$, while the length scale given by $c^2$ over the gravitational acceleration is of the order of $10^{15}\,\rm{m}$. Therefore, the condition (\ref{eq:curvdom}) is fulfilled by four orders of magnitude for the acceleration.

Neglecting quadratic terms in the acceleration, we obtain for the following components of the spacetime metric in the proper detector frame of a non-rotating observer \cite{Ni1978proper} (as above, Latin indices are used for the spatial components with respect to the tetrads and spatial indices are raised and lowered with the spatial metric $\delta_{IJ}=\textrm{diag}(1,1,1)$)
\begin{eqnarray}\label{eq:metricproper}
	\nonumber g^{P}_{00}(c\tau,\mathbf{x})&\approx & -\left(1+\frac{2}{c^2}\mathbf{a}_J(\tau)\mathbf{x}^J + R_{0I0J}(\tau)\mathbf{x}^I \mathbf{x}^J\right)\\
	g^{P}_{0J}(c\tau,\mathbf{\mathbf{x}}) &\approx & - \frac{2}{3}R_{0KJL}(\tau)\mathbf{x}^K \mathbf{x}^L\\
	\nonumber g^{P}_{IJ}(c\tau,\mathbf{x}) &\approx & \delta_{IJ} - \frac{1}{3}R_{IKJL}(\tau) \mathbf{x}^K \mathbf{x}^L\,.
\end{eqnarray}
Since we assumed $s'_\tau(\varsigma)=(0,0,0,1)$ and by construction of the proper detector frame, the proper length of the geodesics $s_\tau(\varsigma)$ is $L_p=b-a$, where the spatial positions of the mirrors are $(0,0,b)$ and $(0,0,a)$ with $b\ge 0$ and $a\le 0$. Then, we find from Eq. (\ref{eq:metricproper}) that $g^{P}_{0z}(c\tau,0,0,z) \approx 0$ for all $\tau$ and $z$ along the resonator. Furthermore, by construction, all segments of the rod remain at fixed coordinate positions along the $z$-axis and we find that $\dot\gamma_\varsigma(\varrho)^\mathcal{M}=((g^{P}_{00})^{-1/2},0,0,0)$. Since $s'_\tau(\varsigma)=(0,0,0,1)$, we obtain $g^{P}_{\mathcal{M}\mathcal{N}}\dot\gamma_\varsigma(\varrho)^\mathcal{M} s^{\prime\mathcal{N}}_\varrho(\varsigma) = (g^{P}_{00})^{-1/2}g^{P}_{0 z}(c\tau(\varrho),0,0,z(\varsigma))$.
From Eq. (\ref{eq:metricproper}) and one of the symmetries of the Riemann tensor $R_\mathcal{MNKL}=-R_\mathcal{MNLK}$ follows that the condition $g^{P}_{\mathcal{M}\mathcal{N}}\dot\gamma_\varsigma(\varrho)^\mathcal{M} s^{\prime\mathcal{N}}_\varrho(\varsigma)=0$, which we assumed in our definition of a rigid resonator in Sec. \ref{sec:rigidrod}, is approximately fulfilled for a small proper length of the resonator \footnote{Here small proper length means that the proper detector frame metric (\ref{eq:metricproper}) is still a valid approximation to the actual spacetime metric.}.

To obtain the frequency of the rigid resonator measured by an observer at $\mathbf{x}$ using Eq. (\ref{eq:fundfreq}), we have to calculate the corresponding radar distance between the mirrors. The radar distance is obtained from the trajectories $\xi_\pm(\iota)$ of light like particles bouncing back and forth between the mirrors as described in Sec. \ref{sec:fundfreq} and illustrated in Fig. \ref{fig:pathlight}. In Sec. \ref{sec:fundfreq}, we already assumed that acceleration and curvature only change very slowly with $\tau$. Under this assumption, we can replace acceleration and curvature in Eq. (\ref{eq:metricproper}) by their values at $\tau_0$. The trajectories $\xi_\pm(\iota)$ have to fulfill the null condition $g^P_\mathcal{MN}(\xi(\iota))\dot\xi^\mathcal{M}(\iota)\dot\xi^\mathcal{N}(\iota)=0$ and the geodesic equation that governs the motion of test particles $\ddot\xi^\mathcal{A}(\iota)=	-\Gamma^\mathcal{A}_\mathcal{BC}(\xi(\iota))\dot\xi^\mathcal{B}(\iota)\dot\xi^\mathcal{C}(\iota)$. In first order in $h^{P}_{\mathcal{M}\mathcal{N}}$, one finds for the Christoffel symbols
\begin{eqnarray}\label{eq:christh}
	\Gamma^\mathcal{A}_{\mathcal{B}\mathcal{C}}	&=& 	\frac{1}{2}\eta^{\mathcal{A}\mathcal{R}}\left(\partial_\mathcal{B} h^{P}_{\mathcal{C}\mathcal{R}}+\partial_\mathcal{C} h^{P}_{\mathcal{B}\mathcal{R}}
		-\partial_\mathcal{R} h^{P}_{\mathcal{B}\mathcal{C}}\right)\,,
\end{eqnarray}
which shows that the Christoffel symbols are of the same order as $h^\rm{P}_\mathcal{MN}$. Then, to first order in $h^\rm{P}_\mathcal{MN}$, the trajectories are given by $\xi^\mathcal{M}_\pm(\iota)=c(\iota_{0,\pm} + \iota,0,0,\pm \iota) + \delta^\mathcal{M}_\pm(\iota)$, where $\iota_{0,\pm}$ are constants and the functions $\delta^\mathcal{M}_\pm(\iota)$ are of the same order as $h^\rm{P}_\mathcal{MN}$. With $R_\mathcal{MNKL}=-R_\mathcal{MNLK}$, we find that $g^P_{zz}\approx 1$ and $g^P_{0z}=g^P_{z0}\approx 0$ along $\xi_\pm(\iota)$, and we obtain that $\dot\delta_\pm^0(\iota) \approx c h^P_{00}(c\iota_{0,\pm},0,0,\pm c\iota)$ and $\dot\delta_\pm^z(\iota) \approx \pm c h^P_{00}(c\iota_{0,\pm},0,0,\pm c\iota)/2$ solve the light cone condition and the geodesic equation.
The difference in coordinate time $\tau$ between sending and receiving the light pulse is given as
\begin{eqnarray}
	\delta \tau &=& \int_{\iota_{+,a}}^{\iota_{+,b}} \dot\xi_+^0(\iota) d\iota  + \int_{\iota_{-,b}}^{\iota_{-,a}} \dot\xi_-^0(\iota) d\iota\,,
\end{eqnarray}
where $\iota_{\pm,a}$ and $\iota_{\pm,b}$ are the parameter values at which the ray intersects with the world lines of mirror A and mirror B, respectively. A transformation of the integration variable to $z_\pm = \xi^z_\pm(\iota)$ leads to
\begin{eqnarray}
	\delta \tau & = & \int_{a}^{b} \frac{\dot\xi_+^0(\iota(z_+))}{\dot\xi_+^z(\iota(z_+))} dz_+  + \int_{b}^{a} \frac{\dot\xi_-^0(\iota(z_-))}{\dot\xi_-^z(\iota(z_-))} dz_- \\
	& \approx & \int_{a}^{b} \frac{c + \dot\delta_+^0(z_+/c)}{c + \dot\delta_+^z(z_+/c)} dz_+  + \int_{b}^{a} \frac{c + \dot\delta_-^0(-z_-/c)}{-c + \dot\delta_-^z(-z_-/c)} dz_- \,,
\end{eqnarray}
which reduces to
\begin{eqnarray}\label{eq:roundtriptime}
	\nonumber \delta \tau &\approx& 2\int_{a}^{b} dz_\pm \left( 1 + \frac{h^\rm{P}_{00}(c\tau_0,0,0,z_\pm)}{2} \right) \\
	&\approx& 2\int_{a}^{b} dz_\pm \left( 1 - \frac{1}{2}\left(\frac{2}{c^2}\mathbf{a}^z(\tau_0) z_\pm + R_{0z0z} (\tau_0) z_\pm^2\right)\right) \\
	\nonumber &\approx & \frac{2}{c} L_p \left(1 - \frac{\mathbf{a}^z(\tau_0)}{2c^2}\beta L_p - \frac{R_{0z0z} (\tau_0)}{24}(3\beta^2+1)L_p^2\right)\,,
\end{eqnarray}
where we defined $\beta:=2b/L_p - 1$ and used $a=(\beta - 1)L_p/2$. Under the assumption of slowly changing acceleration and curvature, the coordinate time $\delta \tau$ needed for a round trip of a light pulse inside the resonator is independent of the point on the $z$-axis where it was sent from and received at, as long as it is sent and recieved at the same point. Therefore, we can calculate the radar length of the resonator measured at a given position $z_0=(\sigma+\beta)L_p/2$ along the $z$-axis inside the resonator ($\sigma\in[-1,1]$) as
\begin{eqnarray}\label{eq:radarlengthrigid}
	 \nonumber R_\sigma &\approx& \sqrt{-g^P_{00}((\tau_0,0,0,z_0))}\frac{c}{2}\delta \tau \\
	 &=&  L_p \left[ 1+\frac{\mathbf{a}^z(\tau_0)}{2c^2}\sigma L_p  + \frac{R_{0z0z} (\tau_0)}{24}\left(3\sigma^2 + 6\sigma\beta - 1\right)L_p^2\right]\,.
\end{eqnarray}
Eq. (\ref{eq:radarlengthrigid}) was calculated for a given time $\tau_0$ to make our assumption of slow changes of acceleration and curvature explicit. Of course, we are free to choose the value of $\tau_0$. Therefore, we can replace $\tau_0$ in Eq. (\ref{eq:radarlengthrigid}) with $\tau$. Then, the relative change of the resonance frequencies measured at $z_0=(\sigma+\beta)L_p/2$ is given as
\begin{eqnarray}\label{eq:fundfreqrigid}
	 \delta_{\omega,\sigma}:=\frac{\omega_{n}}{\bar\omega_n} - 1 &\approx& - \frac{\mathbf{a}^z(\tau)}{2c^2}\sigma L_p  - \frac{R_{0z0z} (\tau)}{24}\left(3\sigma^2 + 6\sigma\beta - 1\right)L_p^2\,,
\end{eqnarray}
where $\bar\omega_n$ is the $n$-th resonance frequency of the resonator for vanishing acceleration and curvature. 


We find that the only linear contribution of the acceleration $\mathbf{a}^z$ to the resonance frequency spectrum in Eq. (\ref{eq:fundfreqrigid}) is via a position-dependent red shift. It vanishes for $\sigma=0$, which corresponds to a frequency measurement in the center of the resonator.
The term $3\sigma^2$ corresponds to a pure red shift with respect to the center of the cavity. The term $6\beta\sigma$ is due to the displacement of the resonator's support from its center. In order to move the support along the trajectory $\gamma(\tau)$, while keeping the proper length of the resonator constant, the acceleration $\mathbf{a}^z_\rm{cm}(\tau)=\mathbf{a}^z(\tau) + c^2R_{0z0z}(\tau)\beta L_p/2$ must be applied to the center of mass of the resonator\footnote{This result can be directly obtained by considering the differential acceleration between the support and the center of the cavity by use of the geodesic deviation equation.}. Based on these considerations, we can rewrite Eq. (\ref{eq:fundfreqrigid}) as
\begin{eqnarray}\label{eq:fundfreqrigidcm}
	 \delta_{\omega,\sigma}&\approx& - \frac{\mathbf{a}^z_\rm{cm}(\tau)}{2c^2}\sigma L_p - \frac{R_{0z0z} (\tau)}{24}\left(3\sigma^2 - 1\right)L_p^2\,.
\end{eqnarray}
However, a realistic rod can never be rigid. In the next section, we will consider the first order deviations from the rigid rod by taking the deformation of the rod due to small inertial and gravitational forces into account.


\section{Deformable optical resonators}
\label{sec:realrod}

In the proper detector frame, every segment of the rod has a world-line with constant spatial components. The acceleration of a segment of the rod at $\mathbf{x}=(c\tau,0,0,z)$, in comparison to a freely falling test particle initially at rest at the same position as that segment, can be derived from the geodesic equation
\begin{eqnarray}\label{eq:geodesicrest}
	\ddot\gamma_{\rm{rest},\mathbf{x}}^\mathcal{M}(\tau_{\rm{rest}})
	&=&	-\Gamma^\mathcal{M}_{\mathcal{A}\mathcal{B}}(\gamma_{\rm{rest},\mathbf{x}})\dot\gamma_{\rm{rest},\mathbf{x}}^\mathcal{A}(\tau_{\rm{rest}})\dot\gamma_{\rm{rest},\mathbf{x}}^\mathcal{B}(\tau_{\rm{rest}})\,,
\end{eqnarray}
where, in first order in the metric perturbation, the tangent for a test particle at rest is $\dot\gamma_{\rm{rest},\mathbf{x}}=(c(-g^P_{00}(\mathbf{x}))^{-1/2},0,0,0)$ with $(-g^P_{00}(\mathbf{x}))^{-1/2}\approx 1 + h^\rm{P}_{00}(\tau,0,0,z)/2$. The dot means the derivative with respect to the curve parameter $\tau_{\rm{rest}}$. In first order in $h^{P}_{\mathcal{M}\mathcal{N}}$, the Christoffel symbols are given by Eq. (\ref{eq:christh}) and are proportional to the metric perturbation. Therefore, expanding Eq. (\ref{eq:geodesicrest}) in first order in the metric perturbation, we find $\ddot\gamma_{\rm{rest},\mathbf{x}}^\mathcal{M}\approx -c^2\Gamma^\mathcal{M}_{00}$. Since $c d\tau/d\tau_{\rm{rest}} =\dot\gamma_{\rm{rest},\mathbf{x}}^0\approx 1 + h^\rm{P}_{00}(\tau,0,0,z)/2$, we obtain $\mathbf{a}_P^J  \approx -c^2\Gamma^J_{00}$ for the proper and tidal accelerations.

We consider the effect of $\mathbf{a}_P$ on the resonator's end mirrors and the resulting deformation of the rod to be negligible in comparison to the direct effect of $\mathbf{a}_P$ on the rod. Then, we obtain the inertial and tidal forces on the rod by multiplication of $\mathbf{a}_P$ with the mass density $\rho$. These forces give rise to stresses within the rod, represented by the stress tensor $\sigma_{KL}$. For static forces and forces that change very slowly, the stresses are related to the strain via Hooke's law as
\begin{equation}
	\varepsilon_{IJ}=(\mathcal{C}^{-1})_{IJKL} \sigma_{KL}\,,
\end{equation}
where $\mathcal{C}^{-1}$ is the inverse of the stiffness tensor for the material the rod is composed of. From the strain, we can calculate the deformation of the rod by integration along the length of the rod from its center of mass. Since the change of diameter of the rod and its deformations in the $x$-$y$-plane are not of interest for us, we can restrict our considerations to $\varepsilon_{zz}$, $\varepsilon_{xz}$ and $\varepsilon_{yz}$. We assume a constant cross section $A$ of the rod, and we assume that the diameter of the rod is much smaller than its length. The contribution of $\varepsilon_{xz}$ and $\varepsilon_{yz}$ on the length of the rod are of second order in the metric perturbation and can be neglected (see Appendix \ref{sec:deformations})  if
\begin{eqnarray}
	\mathbf{a}_{P,\rm{av}}^z &\gg &  \rm{max}\left\{ L_p^5\langle |\mathbf{a}^{x}_{P\rm{max}}|\rangle^2/c_s^2 w_x^4 , L_p^5\langle |\mathbf{a}^{y}_{P\rm{max}}|\rangle^2/c_s^2 w_y^4 \right\}\,.
\end{eqnarray}
where $\mathbf{a}^{x}_{P\rm{max}}$ and $\mathbf{a}^{y}_{P\rm{max}}$ are the maxima of proper acceleration in the $x$-direction and $y$-direction, respectively, $w_x$ and $w_y$ are the diameters of the rod in the $x$-direction and $y$-direction, respectively, and $\mathbf{a}_{P,\rm{av}}^z$ is the largest of the values given by $\langle\beta|\mathbf{a}^z(\tau)|\rangle$ and $\langle(3\beta^2+1)L_pc^2|R_{0z0z}(\tau)|/6\rangle$, where $\langle \rangle$ denotes the averaging over the observation time (see Appendix \ref{sec:deformations} for the derivation). With these considerations, the tidal accelerations in the proper detector frame in the transversal direction can be neglected if the following conditions hold
\begin{eqnarray}
	\mathbf{a}_{P,\rm{av}}^z &\ge &  \rm{max}\left\{ w_x c^2 \langle |R_{0x0x}|\rangle, w_y c^2 \langle |R_{0y0y}|\rangle\right\}\,,
\end{eqnarray}
Additionally, we assume that the various contributions to the transversal tidal acceleration do not oscillate on resonance with any elastic mode of the rod that is not already on resonance with the oscillations of the longitudinal acceleration and the longitudinal tidal acceleration. In most situations of interest, it should be easy to fulfill these conditions by choosing an appropriate orientation of the resonator and appropriate values for $w_x$ and $w_y$. In particular, the conditions are fulfilled for the examples given in Sec. \ref{sec:exampleunif},  \ref{sec:examplebh} and \ref{sec:examplemass}.

Under the above conditions, the only non-zero component of the stress tensor of interest for us is $\sigma_{zz}$ and its relation to the strain is given as
\begin{equation}
	\varepsilon_{zz} = \frac{1}{Y}\sigma_{zz}\,.
\end{equation}
where $Y$ is the Young's modulus of the rod material.
If we assume a constant mass density, the force along the rod in the positive $z$-direction can be obtained as
\begin{eqnarray}\label{eq:posporce}
	F^z_+(\tau,z)
	&=&	\int_z^b dz' \rho A \,\mathbf{a}_P^z(z',\tau) \simeq -\rho A \left(b-z\right) \left(\mathbf{a}^z(\tau)+\frac{c^2}{2}\left(b+z\right)R_{0z0z}(\tau)\right)\,.
\end{eqnarray}
where we made use of $\mathbf{a}^z_P(z,\tau) \simeq -c^2\Gamma^z_{00}(z,\tau) = -(\mathbf{a}^z(\tau) + c^2 R_{0z0z}(\tau)z)$. For the force along the rod in the negative $z$-direction, we find
\begin{eqnarray}\label{eq:negforce}
	F^z_-(\tau,z)
	&\simeq&-\rho A \left(z-a\right)\left(\mathbf{a}^z(\tau)+\frac{c^2}{2}\left(z+a\right)R_{0z0z}(\tau)\right)\,.
\end{eqnarray}
Since the support of the resonator is inside the resonator, we obtain the total deformation of the resonator by integrating the strains $\varepsilon^+_{zz}=F^z_{+}/A$ and $\varepsilon^-_{zz}=F^z_{-}/A$ on the two sides of the resonator from $z=0$ to the ends, respectively. The effective change of the proper length is
\begin{equation}\label{eq:length}
	\begin{array}{lcl}	
	\delta L_p 
	&\simeq&   \int_0^b dz' \varepsilon^+_{zz}(z') + \int^0_a dz' \varepsilon^-_{zz}(z')= -\frac{c^2}{c_s^2} L_p \left(\frac{\mathbf{a}^z(\tau)}{2c^2}\beta L_p + \frac{R_{0z0z}(\tau)}{12}(3\beta^2+1)L_p^2\right)\,.
	\end{array}
\end{equation}
where $c_s=\sqrt{Y/\rho}$ is the speed of sound in the rod material. The acceleration induces a contraction of one side of the resonator and an expansion of the other. Therefore, the acceleration amounts to a change of the proper length, proportional to the displacement $\beta L_p/2$ of the support with respect to the center of the resonator. The change of the proper length proportional to $R_{0z0z}(\tau)$ can be split into two terms. The term proportional to $\beta^2$ corresponds to the acceleration $\mathbf{a}^z_\rm{cm}(\tau_0)=\mathbf{a}^z(\tau_0) + c^2R_{0z0z}(\tau)\beta L_p/2$ of the center of mass of the resonator that we discussed at the end of Sec. \ref{sec:fundfreqrigid}. For a freely falling resonator ($\beta=0=\mathbf{a}^z(\tau)$), only the second term in the brackets remains.

From Eq. (\ref{eq:fundfreqrigid}) and Eq. (\ref{eq:length}), we find for the relative change of the resonance frequencies of the deformable resonator
\begin{eqnarray}\label{eq:fundfreqdefo}
	  \nonumber \delta_{\omega,\sigma} &\approx & - \frac{\delta L_p}{L_p} - \left(\frac{\mathbf{a}^z(\tau)}{2c^2}\sigma L_p  + \frac{R_{0z0z}(\tau)}{24}\left(3\sigma^2 + 6\sigma\beta - 1\right)L_p^2\right)	  \\ 	 & \approx & \frac{\mathbf{a}^z(\tau)}{2c^2}\left(\frac{c^2}{c_s^2}\beta - \sigma\right)L_p  + \frac{R_{0z0z}(\tau)}{24}\left(2\frac{c^2}{c_s^2}(3\beta^2+1) - 3\sigma^2 - 6\sigma\beta + 1 \right)L_p^2
\,.
\end{eqnarray}
Note that the deformation of the resonator changes the coordinate position of every point inside the resonator \footnote{ Any deformation of the rod also leads to a change of density and the speed of sound in the rod which, in turn, leads to a modulation of the deformation of the rod. We consider this effect to be negligible here. In particular, it corresponds to a non-linear correction of Hook's law. Therefore, the result in Eq. (\ref{eq:fundfreqdefo}) can be considered accurate as long as Hook's law can be applied. As the deformations considered are supposed to be small, Hook's law should hold with a very good accuracy. }. This leads to a change in the trajectory of a light pulse within the resonator, and the whole calculation we made in Sec. \ref{sec:fundfreqrigid} would be changed. However, this change would only amount to a change of the resonance frequencies in second order in the metric perturbation and we can neglect it. 

Again, we can write the relative shift of the resonance frequencies in a neater way using the center of mass acceleration as
\begin{eqnarray}\label{eq:fundfreqdefocm}
	   \delta_{\omega,\sigma} &\approx & \frac{\mathbf{a}^x(\tau)_\rm{cm}}{2c^2}\left(\frac{c^2}{c_s^2}\beta -  \sigma \right)L_p  + \frac{R_{0z0z}(\tau)}{24}\left(2\frac{c^2}{c_s^2} + 1 - 3\sigma^2 \right)L_p^2\,.
\end{eqnarray}
As expected, we would obtain the result in Eq. (\ref{eq:fundfreqrigidcm}) for the Born rigid rod from Eq. (\ref{eq:fundfreqdefocm}) if the speed of sound in the material was infinite. This coincides with the observation that a Born rigid rod violates causality, as its segments would need to interact with an infinite speed. A more realistic definition of a rigid rod was given in \cite{Natario:2014koa} as a rod in which the speed of sound is equivalent to the speed of light. In Appendix \ref{sec:causal}, we show that the approach of \cite{Natario:2014koa} leads to the same expression of the change of the length of the rigid rod as our Eq. (\ref{eq:length}). The relative shift of the resonance frequencies for such a causal rigid rod is found from Eq. (\ref{eq:fundfreqdefocm}) in the limit $c_s \rightarrow c$ as
\begin{eqnarray}\label{eq:fundfreqdefocmcaus}
	   \delta_{\omega,\sigma} &\approx & \frac{\mathbf{a}^z(\tau)_\rm{cm}}{2c^2}\left(\beta -  \sigma \right)L_p  + \frac{R_{0z0z}(\tau)}{8}\left(1 - \sigma^2 \right)L_p^2\,.
\end{eqnarray}
In particular, we find that the contribution of curvature to the relative frequency shift vanishes if the frequency is measured at one of the mirrors corresponding to $\sigma=\pm 1$. 

However, the speed of sound $c_s$ in every realistic material is always much smaller than the speed of light: for example the speed of sound in aluminum is of the order $5\times 10^3 {\rm m}/{\rm s}$. To date, the material with the highest ratio of Young's modulus and density $Y/\rho=c_s^2$ is carbyne, with a value of the order of $10^9{\rm m}^2/{\rm s}^2$ \cite{liu2013carbyne}, which would correspond to a speed of sound of the order of $3\times 10^4{\rm m}/{\rm s}$. Therefore, we find that the effect of the deformation of matter is by far the most dominant and the rod is far from rigid (may it be Born rigid or causal rigid) in all realistic situations. Finally, we want to point out that the ratio of Young's modulus and density is called the specific modulus. In this sense, $c^2$ can be thought of as the specific modulus of space time. It is interesting to note that this value is off by a factor $4$ from the value $4c^2$ given for the specific modulus of space time in \cite{Tenev2016}.

\section{Deformable dielectric optical resonators}
\label{sec:dielectric}

\begin{figure}[h]
\includegraphics[width=8cm,angle=0]{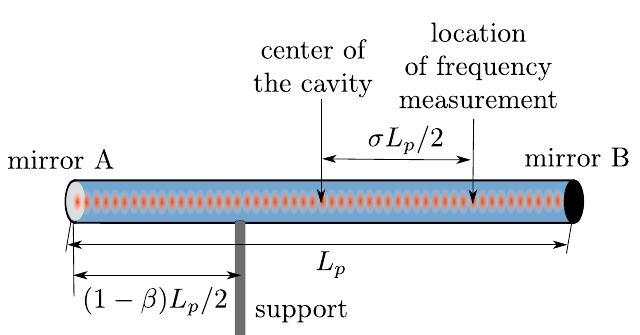}
\caption{\label{fig:rigidcavitydiel} In the case of a dielectric optical resonator, we consider the rod itself to be the resonator.}
\end{figure}

 Up to this point, we have only discussed the case of an empty cavity resonator. Now, let us assume that the rod itself is the optical resonator. In particular, we assume that it consists of an isotropic homogeneous dielectric medium (see Fig. \ref{fig:rigidcavitydiel}). In \cite{Gordon:1923lic}, it was shown that light rays in an isotropic dielectric follow light like geodesics with respect to the dielectric metric tensor (see also \cite{Perlick:2000ray,Hartmann:1992aus})
\begin{equation}\label{eq:gordon}
	g^{P,\rm{diel}}_{\mathcal{M}\mathcal{N}} = g^P_{\mathcal{M}\mathcal{N}} - \left(\frac{c_\rm{diel}^2}{c^2} - 1\right)u_\mathcal{M} u_\mathcal{N}\,,
\end{equation} 
where $c_\rm{diel}^2=(\epsilon\mu)^{-1}$ is the speed of light inside the medium and $u^\mathcal{M}=g^{P\,\mathcal{M}\mathcal{N}}u_\mathcal{N}$ is the normalized tangent vector to the world-line associated with the local segments of the dielectric. 
In our case, these are the segments of the resonator, and therefore, $u^\mathcal{M}(z)=(1 + h^{P}_{00}/2,0,0,0)$ and $u_\mathcal{M}(z)\approx (-1 + h^{P}_{00}/2,h^{P}_{01},h^{P}_{02},h^{P}_{03})$. From Eq. (\ref{eq:gordon}), we obtain the metric
\begin{eqnarray}\label{eq:metricproperdiel}
	\nonumber g^{P,\rm{diel}}_{00}(c\tau,\mathbf{x})&\approx & -\frac{c_\rm{diel}^2}{c^2}\left(1+\frac{2}{c^2}\mathbf{a}_J(\tau)\mathbf{x}^J + R_{0I0J}(\tau)\mathbf{x}^I \mathbf{x}^J\right)\\
	g^{P,\rm{diel}}_{0J}(c\tau,\mathbf{\mathbf{x}}) &\approx & - \frac{2}{3}\frac{c_\rm{diel}^2}{c^2} R_{0KJL}(\tau)\mathbf{x}^K \mathbf{x}^L\\
	\nonumber g^{P,\rm{diel}}_{IJ}(c\tau,\mathbf{x}) &\approx & \delta_{IJ} - \frac{1}{3}R_{IKJL}(\tau) \mathbf{x}^K \mathbf{x}^L\,.
\end{eqnarray}
Now, all of the considerations made for the empty resonator above can also be made for a resonator composed of an isotropic, homogeneous dielectric by using the metric $g^{P,\rm{diel}}_{\mathcal{M}\mathcal{N}}$ for the propagation of the phase fronts given by the eikonal function. Hence, we obtain the resonance frequencies in an isotropic homogeneous dielectric by multiplying the result for the empty resonator with $c_\rm{diel}/c$. This factor cancels in the relative frequency perturbation so that
\begin{equation}
\delta_{\omega,\sigma}^\rm{diel} = \delta_{\omega,\sigma}\,.
\end{equation} 
A similar metric as in (\ref{eq:gordon}) has been shown to arise for particles or quasi particles in other matter systems, e.g. for electrons in graphene \cite{Zubkov:2013sja}. Our analysis may also apply to these situations.

\section{Example: Uniform acceleration}
\label{sec:exampleunif}

To illustrate the applicability of our results, we will consider some examples in the following. A particularly straightforward example is the situation of a non-rotating resonator that is uniformly accelerated along the optical axis. From the equivalence principle follows that this situation is similar to the situation of an optical resonator kept vertically at a fixed position in the gravitational field of a massive object like the earth. However, since we are considering an extended object, the curvature of the gravitational field would also enter the frequency spectrum of the resonator as in Eq. (\ref{eq:fundfreqdefocm}). Hence, the effect of uniform acceleration and a gravitational field do only coincide if the effect of curvature can be neglected. For uniform acceleration, we find
\begin{eqnarray}\label{eq:fundfreqdefounif}
	   \delta_{\omega,\sigma} &\approx & \left(\frac{\beta}{c_s^2} -  \frac{\sigma}{c^2} \right)\frac{\mathbf{a}^xL_p}{2}  \,.
\end{eqnarray}
For $\beta=\pm 1$, a length of the resonator of $L_p\sim 2\,\rm{cm}$, an acceleration of the order of $10\,\rm{ms^{-2}}$, which is similar to the gravitational acceleration of the earth, and a speed of sound in the rod of the order of $10^3\,\rm{ms^{-1}}$ (similar to the speed of sound in aluminum), we obtain a relative frequency shift of the order of $10^{-7}$. This frequency shift is given only by the first term in Eq. (\ref{eq:fundfreqdefounif}) as the second term is smaller by about 11 orders of magnitude. Since the first term is due to the deformation of the resonator it is a Newtonian effect.

For the case $\beta=0$ the first term in (\ref{eq:fundfreqdefounif}) vanishes. What remains is a purely relativistic effect due to a difference in proper time between the center of the resonator and every other point along the optical axis. Setting the parameter $\sigma$ to $-1$ and $+1$ means that the frequency is measured at the mirror A and mirror B, respectively. We find a relative frequency shift of the order of $\mp 10^{-18}$. The measurement of such a small frequency shift seems to be experimentally challenging but may be feasible with state of the art technology. For example, currently, optical clocks reach a relative precision of $10^{-18}$ over an integration time of $1\,\rm{s}$ \cite{Hinkley:2013ana,Ushijima:2015cry}. Of course, higher frequency shifts can be reached with longer cavities and larger accelerations.

\section{Example: Plunge into a black hole}
\label{sec:examplebh}

\begin{figure}[h]
\includegraphics[width=7cm,angle=0]{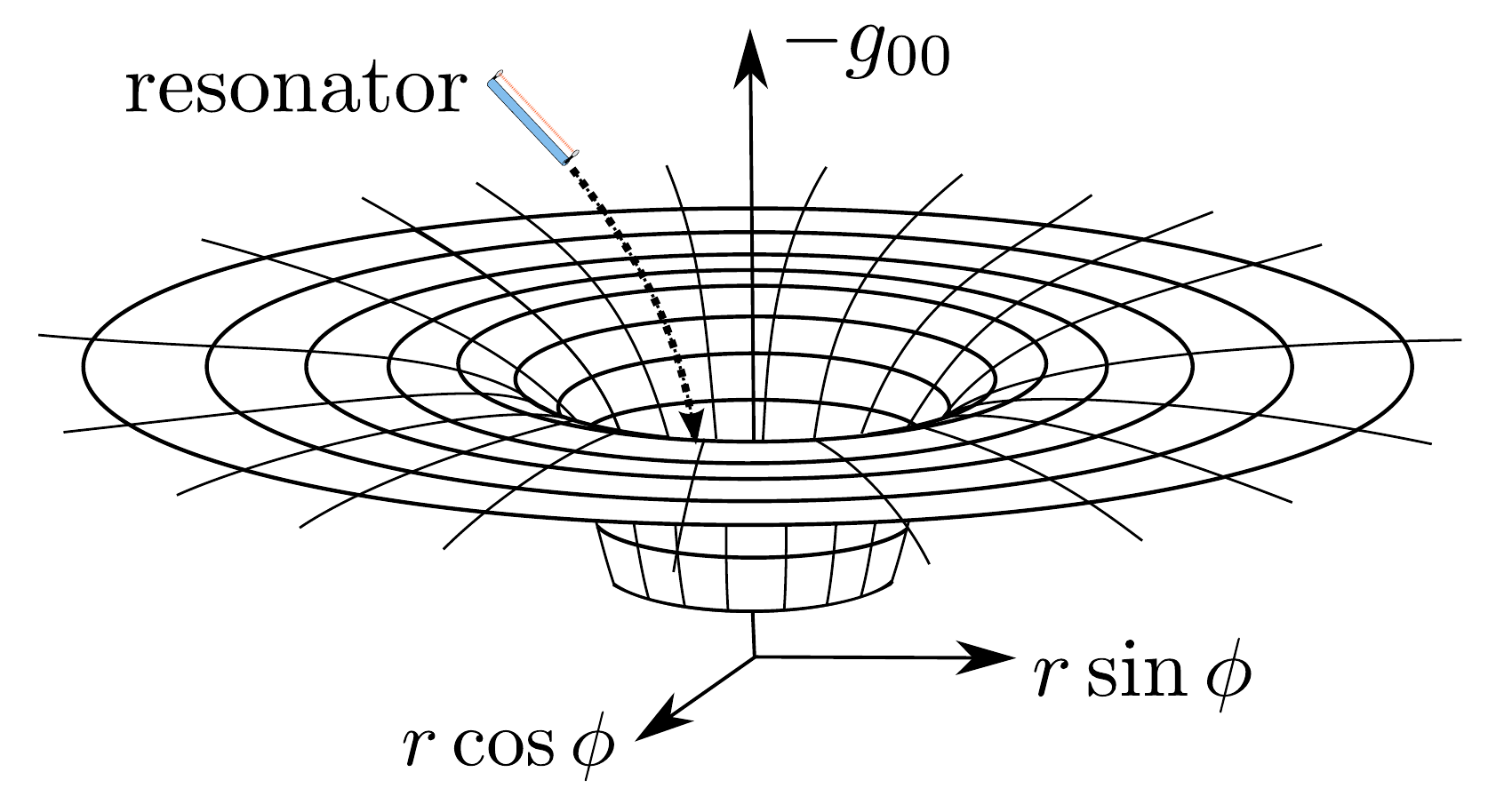}
\caption{\label{fig:bh} Artistic representation of the optical resonator plunging radially into a black hole.}
\end{figure}

We consider the results derived in this article as a basis for optomechanics in relativity and gravity which implies their application to experiments in laboratories on the surface of the Earth or in space. However, our approach is not limited to spacetimes that only bear weak gravitational effects. It is the spacetime metric seen by the optical resonator in its proper detector frame that has to be a linearized metric. This is ensured by the condition $l_\rm{var}\gg L_p$.
To illustrate the applicability of our results to spacetimes with strong gravitational effects, we consider the situation of a non-rotating resonator that falls into a non-rotating black hole (see Fig. \ref{fig:bh}). 
To this end, we consider the Schwarzschild metric in spherical Schwarzschild coordinates $(ct,r,\vartheta,\phi)$
\begin{equation} \label{eq:schwarzschild}
	g = \text{diag} \left(-f(r), \frac{1}{f(r)}, r^2, r^2 \sin^2(\vartheta) \right)\,, 
\end{equation}
where $f(r) = 1-r_S/r$ and $r_S$ is the Schwarzschild radius. We assume that the support of the resonator falls radially from $r=R$ into the center of the black hole at $\phi=0$ and $\vartheta=\pi/2$. The corresponding trajectory is given in \cite{Misner1973} as
\begin{eqnarray}
	 \label{eq:trajectoryr} r(\varrho) &=&R\cos^2(\varrho/2)\\
	\label{eq:trajectorytau} c\tau(\varrho) &=&\frac{R}{2}\left(\frac{R}{r_S}\right)^{1/2}\left(\varrho + \sin\varrho\right)\,,
\end{eqnarray}
parameterized by $\varrho$. We see that $r=0$ for $\varrho=\pi$, which means that the singularity at the center of the black hole is reached in finite proper time $\tau=\pi R^{3/2}/2cr_S^{1/2}$. The tangent to the world line of the falling support of the resonator is
\begin{equation}
	\dot\gamma^\mu=c\left(\frac{\sqrt{f(R)}}{f(r(\varrho))},-\sqrt{\frac{r_S}{R}}\frac{\sin\varrho}{1+\cos\varrho},0,0\right)\,
\end{equation}
where $\varrho=\varrho(\tau)$ is implicitly given by Eq. (\ref{eq:trajectorytau}), $\dot\gamma^1$ can be obtained directly from Eq. (\ref{eq:trajectoryr}) and Eq. (\ref{eq:trajectorytau}) and $\dot\gamma^0$ can be found from the normalization condition $\dot\gamma^\mu\dot\gamma^\nu g_{\mu\nu}(r(\varrho))=-c^2$. Then, the time line can be found as $\gamma =(ct(\tau),r(\varrho(\tau)),\pi/2,0)$, where $ct(\tau)=\int^\tau_0 d\tau'\,\dot\gamma^0(\varrho(\tau'))$. An orthonormal tetrad that is parallel transported along the time like geodesic $\gamma$ is given as
\begin{eqnarray}\label{eq:tetradss}
	\nonumber \tilde\epsilon^\mu_0 &=&\dot\gamma^\mu/c\,,\\
	\nonumber \tilde\epsilon^\mu_1 &=&\left(-\sqrt{\frac{r_S}{R}}\frac{\tan(\varrho/2)}{f(r(\varrho))},\sqrt{f(R)},0,0\right)\,,\\
	\tilde\epsilon^\mu_2 &=&(0,0,r(\varrho)^{-1},0)\quad\rm{and}\\
	\nonumber \tilde\epsilon^\mu_3 &=&(0,0,0,r(\varrho)^{-1})\,.
\end{eqnarray}
All other orthonormal tetrads can be obtained by orthogonal transformations in three dimensions on the spatial part of the tetrad (\ref{eq:tetradss}). Due to the spherical symmetry of the spacetime and the radial trajectory of the resonator at $\vartheta=\pi/2$ and $\phi=0$, we can restrict our considerations to rotations in the $\epsilon^\mu_1$-$\epsilon^\mu_3$-plane. Then, we define the rotated frame
\begin{eqnarray}\label{eq:tetradssfin}
	\nonumber \epsilon^\mu_0 &=& \tilde\epsilon^\mu_0\,,
	\quad \epsilon^\mu_1 =\cos\varphi\, \tilde\epsilon^\mu_1 + \sin\varphi\, \tilde\epsilon^\mu_3\,,\\
	\tilde\epsilon^\mu_2 &=& \tilde\epsilon^\mu_2 \quad\rm{and}\quad 	\epsilon^\mu_3 = \cos\varphi\, \tilde\epsilon^\mu_3 - \sin\varphi\, \tilde\epsilon^\mu_1\,,
\end{eqnarray}
where the angle $\varphi\in [0,\pi/2]$ gives the orientation of the resonator in the $\epsilon^\mu_1$-$\epsilon^\mu_3$-plane. From the tetrad (\ref{eq:tetradssfin}), we obtain the proper detector frame. The $z$-direction is defined by $\epsilon^\mu_3$ and we find from Eq. (\ref{eq:fundfreqdefo}) that 
\begin{eqnarray}
	  \nonumber \delta_{\omega,\sigma} & \approx & \frac{R_{0z0z}(\tau)}{24}\left(2\frac{c^2}{c_s^2}(3\beta^2+1) - 3\sigma^2 - 6\sigma\beta + 1 \right)L_p^2
\,,
\end{eqnarray}
where no proper acceleration appears since the resonator is assumed to be freely falling. The curvature tensor component $R_{0z0z}(\tau)$ is explicitly given as
\begin{eqnarray}\label{eq:curvzSS}
	R_{0z0z}(\tau)&=&\epsilon^\mu_0 \epsilon^\nu_3 \epsilon^\rho_0 \epsilon^\sigma_3 R_{\mu\nu\rho\sigma}(r(\varrho)))\\	
	\nonumber &=& \cos^2\varphi\, \left(\frac{f(R)}{f(r(\varrho))}\right)^2 R_{\bar{0}r\bar{0}r}(r(\varrho)) \\
	&& + \frac{\sin^2\varphi}{R^2}\Bigg(\frac{f(R)\sec^4(\varrho/2)}{f(r(\varrho))^2} R_{\bar{0}\phi\bar{0}\phi}(r(\varrho)) 
	+ \frac{r_S}{R}\frac{4\tan^2(\varrho/2)}{(1+\cos\varrho)^2} R_{r \phi r \phi}(r(\varrho))\Bigg)\,.
\end{eqnarray}
Here, we used that $R_{\bar{0} \phi r \phi}=0$ for the Schwarzschild metric. The expressions for the other curvature tensor components appearing in Eq. (\ref{eq:curvzSS}) at $\vartheta=\pi/2$ are given as
\begin{eqnarray}
	\nonumber R_{\bar{0}r\bar{0}r}(r)=-\frac{r_S}{r^3}\,,\quad R_{\bar{0}\phi\bar{0}\phi}(r)=f(r)\frac{r_S}{2r}\\
	\rm{and}\quad R_{r \phi r \phi}(r)=-f(r)^{-1}\frac{r_S}{2r}\,.
\end{eqnarray}
We obtain 
\begin{eqnarray}\label{eq:curvcavss}
	R_{0z0z}(\tau)&=& -\frac{(1+3\cos(2\varphi))r_S}{4r(\varrho)^3}\,,
\end{eqnarray}
and we find that the specification of the angle of orientation of the rod $\varphi$ gives rise to a numerical factor which vanishes only at $\varphi=\arccos(-1/3)/2$. Hence, for $\varphi\neq\arccos(-1/3)/2$, the frequency shift is proportional to the frequency shift at $\varphi=0$, which corresponds to vertical orientation. 
\begin{figure}[h] 
\includegraphics[width=10cm,angle=0]{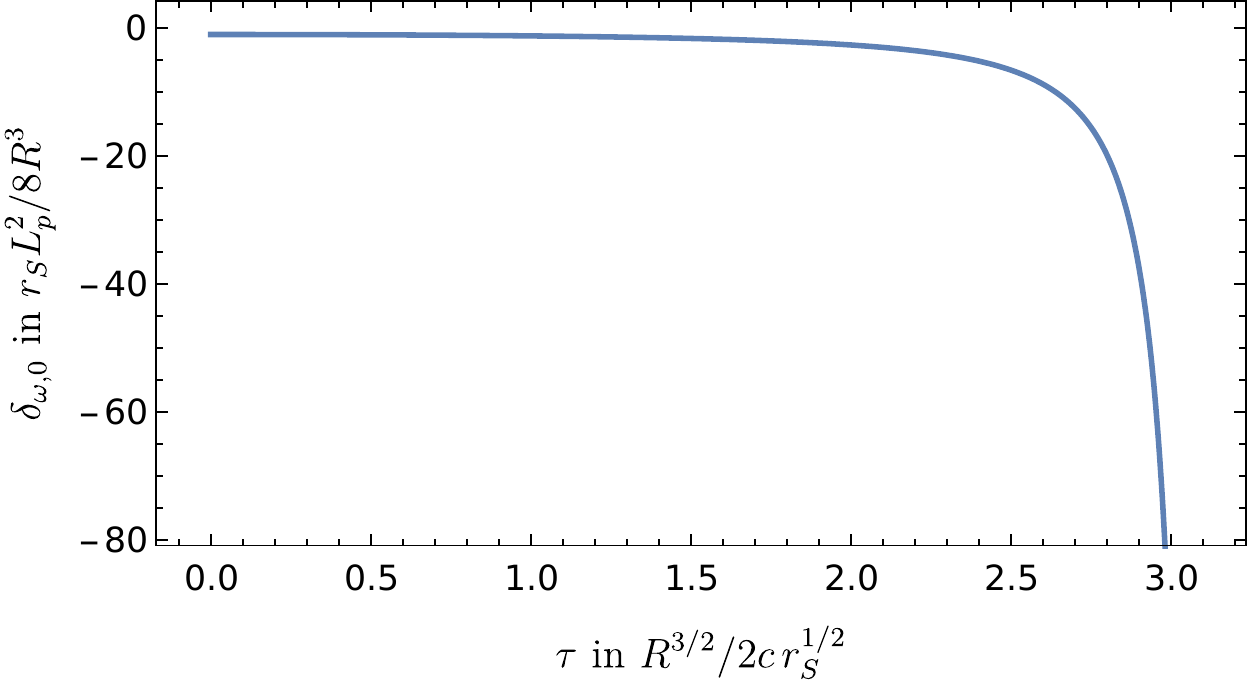}
\caption{\label{fig:example_schwarzschild} The frequency shift of a vertically oriented optical resonator falling into a black hole is plotted over the normalized proper time measured at the center of the resonator. }
\end{figure}
For a vertically oriented causal rigid resonator supported at its center, we find the relative frequency shift at its center is given by
\begin{eqnarray}\label{eq:shiftSS}
	  \delta_{\omega,0}(\tau) & \approx & -\frac{r_SL_p^2}{8r(\varrho)^3}
\,.
\end{eqnarray}
The time evolution of this frequency shift is plotted in Fig. \ref{fig:example_schwarzschild}. We see that the frequency shift in Eq. (\ref{eq:shiftSS}) stays finite until $r=0$ is reached at $\varrho(\tau)=\pi$. In particular, there is no effect due to the crossing of the event horizon at $r_S$. As stated at the beginning of this section, our approach is accurate only for $l_\rm{var}\gg L_p$. From Eq. (\ref{eq:curvcavss}), we find that $l_\rm{var}=\sqrt{r(\varrho)^3/r_S}$ for $\varphi=0$. The stellar black hole has a Schwarzschild radius of the order of $10^3\,\rm{m}$. For an optical resonator of a length of the order of $10^{-2}\,\rm{m}$, this implies that that our approach breaks down when a radius of the order of $1\,\rm{m}$ is reached which is far beyond the event horizon at $r=r_S$.

The effect of the event horizon can be seen by considering a situation in which the measured frequency is imprinted on a signal at the center of the resonator and sent out radially to an observer that stays at constant coordinate $r=R>r_S$. This observer receives a signal with frequency
\begin{eqnarray}\label{eq:omeganR}
	  \nonumber \omega_{n,R}(t) & \approx &  \sqrt{\frac{f(r(t))}{f(R)}}\left(\sqrt{f(R)}+\sqrt{\frac{r_S}{R}}\frac{\sin\varrho(\tau(t))}{1+\cos\varrho(\tau(t))}\right)^{-1}\sqrt{f(r(t))}\,\frac{c n\pi}{L_p}\left(1+\delta_{\omega,0}(\tau(t))\right)
\,,
\end{eqnarray}
where $r(t)$ and $\tau(t)$ are given implicitly by the time line $\gamma(\tau)$. The first factor on the right hand side of Eq. (\ref{eq:omeganR}) corresponds to the gravitational red shift and the second factor to the Doppler shift due to the relative velocity between the emitter and the receiver. The red shift factor $f(r(t))^{1/2}$ vanishes when the resonator passes the event horizon and becomes imaginary.


\section{Example: An oscillating mass}
\label{sec:examplemass}

\begin{figure}[h] 
\includegraphics[width=8cm,angle=0]{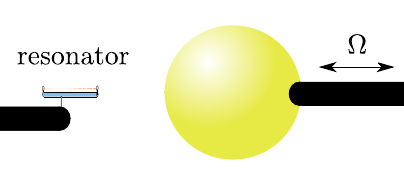}
\caption{\label{fig:resinfrontofgoldmass} Illustration (not to scale) of the resonator placed in front of a gold sphere that oscillates by a lever with frequency $\Omega/2\pi$. The gravitational field of the sphere induces a change of the resonance frequencies of the resonator.}
\end{figure}

As a third example, we consider the situation of a non-rotating resonator in the gravitational field of an oscillating solid sphere of massive matter. The result could be used to consider the possibility of detecting the gravitational field of a small sphere of dense material, like gold or tungsten (see Fig. \ref{fig:resinfrontofgoldmass}). This situation is similar to the one considered in \cite{Schmoele:2017diss} and \cite{Schmoele:2016mde}, where the resonator is a second massive sphere on a support with restoring force. Here we will restrict ourselves to the derivation of the resonance frequency spectrum and an evaluation of its relative change for certain realistic experimental parameters. Also, we assume that the solid sphere is the only source of a gravitational field affecting the optical resonator. To model an earthbound experiment, the gravitational field of the earth would have to be taken into account as well.
To derive our model of the gravitational field of a massive sphere, we start from the Schwarzschild metric, which is given as
\begin{equation}\label{eq:Smetric}
	g^S_{\mu\nu}= (1 + \frac{r_S}{4R})^4\left(\begin{array}{cccc} - \frac{(1 - \frac{r_S}{4R})^2}{(1 + \frac{r_S}{4R})^6} & 0 & 0 & 0  \\
	 0 & 1 & 0 & 0 \\
	 0 & 0 & 1 & 0 \\
	 0 & 0 & 0 & 1 \end{array}   \right)\,,
\end{equation}
in isotropic Cartesian coordinates $(\tilde x^0=c\tilde t,\tilde x,\tilde y,\tilde z)$, where $r_S:=2GM/c^2$ is the Schwarzschild radius of the source mass and $R:=(\tilde x^2+\tilde y^2+\tilde z^2)^{1/2}$. To first order in $r_S/R$, the difference of (\ref{eq:Smetric}) from the Minkowski metric $\rm{diag}(-1,1,1,1)$ has only four non-zero components, namely $h^S_{\tilde{0}\tilde{0}}=h^S_{\tilde x\tilde x}=h^S_{\tilde y\tilde y}=h^S_{\tilde z\tilde z}=\frac{r_S}{R}$. Let us assume that the sphere moves much more slowly than the speed of light and that we are close enough to the sphere so that all changes of the gravitational field can be considered to be instantaneous. With this, we can model the metric perturbation for the moving sphere by replacing $R$ by $R(\tilde t):=
\left((\tilde x-\gamma_M^{\tilde x}(\tilde t))^2+(\tilde y-\gamma_M^{\tilde y}(\tilde t))^2+(\tilde z-\gamma_M^{\tilde z}(\tilde t))^2\right)^{1/2}$, where $\gamma_M^\mu(\tilde t)$ is the trajectory of the source mass. The resulting metric perturbation becomes
\begin{eqnarray}\label{eq:hMpert}
	h^M_{\tilde 0\tilde 0}=h^M_{\tilde x\tilde x}=h^M_{\tilde y\tilde y}=h^M_{\tilde z\tilde z}=\frac{r_S}{R(\tilde t)}=:h^M\\
	h^M_{\mu\nu}=0\,\,\mathrm{for}\,\,\mu\neq\nu\,.
\end{eqnarray}
We assume that the support of the resonator is at rest in the isotropic coordinates on the $\tilde z$ axis in the negative $\tilde z$ direction. To be completely accurate, we would need to fix the proper distance between the support of the resonator and the average position of the sphere, as this corresponds to the assumption that the distance is fixed by another matter system. Furthermore, in every realistic situation, the proper distance would change as the matter system is affected by the gravitational field of the sphere and the gravitational force experienced by the resonator. However, any small error in the position of the resonator will be negligible, as it corresponds to a small change of the acceleration and curvature that we already assumed to be small. From Eq. (\ref{eq:christh}),
we find that an acceleration $a^{\tilde{z}}(\tau)= (\nabla_{\dot\gamma(\tau)}\dot\gamma(\tau))^{z} \approx  c^2\Gamma^{\tilde z}_{00} \approx  -c^2 r_S/2R(\tau)^2$  along the $\tilde z$-axis is necessary to keep the resonator at a fixed position $\tilde z_0 < 0$ on the $\tilde z$-axis, i.e. $\gamma(\tau)=(\tau,0,0,\tilde z_0)$. For the linearly perturbed metric, the curvature tensor is given as
\begin{eqnarray}\label{eq:curvh}
	 R^\alpha_{\;\beta\gamma\delta}
	&\simeq&	\frac{1}{2}\eta^{\alpha	\rho}\left(\partial_\beta\partial_\gamma h^M_{\delta\rho}
		-\partial_\beta\partial_\delta h^M_{\gamma\rho} -\partial_\gamma\partial_\rho h^M_{\beta\delta}
		+\partial_\delta\partial_\rho h^M_{\beta\gamma}\right)\,.	
\end{eqnarray}
We assume that the resonator is fixed along the $\tilde{z}$-axis. From the Eq. (\ref{eq:curvh}), we obtain the curvature component $R_{\tilde{0}\tilde{z}\tilde{0}\tilde{z}}(\tau)=-r_S/R(\tau)^3$. 

To construct the proper detector frame, we need to fix the tetrad corresponding to the observer at the support of the cavity. Since we assume that the support stays at rest in the coordinates $(\tilde x^0,\tilde x,\tilde y,\tilde z)$, we have $\epsilon^\mu_0=((g^S_{\tilde{0}\tilde{0}})^{-1/2},0,0,0)$. We define the three spatial vectors of the tetrad $\epsilon^\mu_J$ with $J=1$, $J=2$ and $J=3$ such that they point in the $\tilde{x}$-direction, $\tilde{y}$-direction and $\tilde{z}$-direction, respectively. Therefore, we find $\epsilon^\mu_1=(0,(g^S_{\tilde{x}\tilde{x}})^{-1/2},0,0)$, $\epsilon^\mu_2=(0,0,(g^S_{\tilde{y}\tilde{y}})^{-1/2},0)$ and $\epsilon^\mu_3=(0,0,0,(g^S_{\tilde{z}\tilde{z}})^{-1/2})$. We conclude that the transformation to the proper detector frame is a linearized coordinate transformation. A linearized coordinate transformation leaves the curvature tensor invariant and we obtain $R_{0z0z}(\tau)=-r_S/R(\tau)^3$. Furthermore, $a^{z}(\tau)=\varepsilon^z_\mu a^\mu(\tau)\approx a^{\tilde{z}}(\tau)$ to first order in the metric perturbation.

Let us assume that the motion of the sphere can be described as $R(\tau)=R_0 + \delta R_0 \sin\Omega\tau$, where $R_0$ is the average distance between the sphere and the position of the support of the resonator, $\delta R_0$ is the amplitude of the sphere's oscillation and $2\pi\Omega$ its frequency. If we assume that $\delta R_0$ is much smaller than $R_0$, the proper acceleration and the curvature can be written as
\begin{eqnarray}\label{eq:aexpansion}
	a^z(\tau) &\approx & -\frac{c^2 r_S}{2R_0^2} \left(1-\frac{2\delta R_0}{R_0}\sin(\Omega \tau +\varphi) \right)\\
	\label{eq:Rexpansion}
	R_{0z0z}(\tau) &\approx & -\frac{r_S}{R_0^3} \left(1 - \frac{3\delta R_0}{R_0}\sin(\Omega \tau +\varphi) \right)\,.
\end{eqnarray}
The first terms in (\ref{eq:aexpansion}) and (\ref{eq:Rexpansion}) are constant, and we can calculate their effect on the frequency spectrum using Eq. (\ref{eq:fundfreqdefo}). 
The resulting time dependent resonance frequencies are given by Eq. (\ref{eq:fundfreqdefo}) as
\begin{eqnarray}\label{eq:exdeltaomega}
	   \delta_{\omega,\sigma} & \approx & -\frac{r_SL_p}{4R_0^2}\left(\left(\frac{c^2}{c_s^2}\beta - \sigma \right) + \left(2\frac{c^2}{c_s^2}(3\beta^2+1)-3\sigma^2 - 6\beta\sigma + 1\right)\frac{L_p}{6R_0}\right)\,.
\end{eqnarray}
Let us assume that the sphere is of gold or tungsten, that the mass of the sphere is $100\,$g (corresponding to a radius of the order of $r_\rm{sph} \sim 1\,\rm{cm}$), which corresponds to a Schwarzschild radius of the order of $10^{-27}\,\rm{m}$, the amplitude of the oscillations $\delta R_0$ is of the order $1\,$mm, while the length of the resonator and $R_\rm{min}$, the minimal distance between the resonator and the sphere, are of the order of $1\,\rm{cm}$. Then, we find that $R_0 =  r_\rm{sph} + \delta R_0 + R_\rm{min} + L_p(1+\beta)/2$ takes values between $2-3\,\rm{cm}$. This results in values for acceleration and spacetime curvature of the order of $10^{-10}\,\rm{ms^{-2}}$ and $10^{-25}\,\rm{m^{-2}}$, respectively. 
We mentioned above that the speed of sound in a rod of aluminum is about $5\times 10^3 {\rm m}/{\rm s}$. Therefore, the relative change of the resonance frequencies of a resonator with its length fixed by an aluminum rod, in the gravitational field of the moving mass, yields $\delta_{\omega} \sim \mp 10^{-18}$ for $\beta=\pm 1$, where the acceleration is dominant, and $\delta_{\omega} \sim 10^{-19}$ for $\beta=0$, where only the curvature contributes. The relativistic effects in Eq. (\ref{eq:exdeltaomega}) are ten orders of magnitude smaller. Hence, to detect them, the whole experimental setup would need to be under control with this precision.

For oscillation frequencies $\Omega$ far below any resonances of the rod holding the optical resonator, we can also derive the effect of the sinusoidally modulated terms in  (\ref{eq:aexpansion}) and (\ref{eq:Rexpansion}) with Eq. (\ref{eq:fundfreqdefo}). We find
\begin{eqnarray}
	   \delta^\Omega_{\omega,\sigma} & \approx & \frac{r_SL_p \delta R_0}{2R_0^3}\sin(\Omega \tau +\varphi)\left(\left(\frac{c^2}{c_s^2}\beta - \sigma \right) + \left(2\frac{c^2}{c_s^2}(3\beta^2+1)-3\sigma^2 - 6\beta\sigma + 1\right)\frac{L_p}{4R_0}\right)\,.
\end{eqnarray}
For the parameters used above, we find for $\beta=\pm 1$ an amplitude of the frequency oscillations of the order of $10^{-19}$. The temporal modulation of the frequency shift may be an advantage in experimental situation as it may be used to increase sensitivity.  As for the example of uniform acceleration, the values for the frequency shifts that we found for this setup seem to be challenging but not out of reach of state of the art experimental techniques.


\section{Conclusions and Outlook}
\label{sec:conclusions}

We derived an expression for the resonance frequencies of an optical resonator moving in a weak gravitational field in a relativistic setup. Firstly, we considered a Born-rigid resonator, which we assumed to be constructed from a Born-rigid rod. Secondly, we considered a deformable resonator, where we assumed the rod to consist of a realistic material with finite Young's modulus. In this context, we discussed the concept of a causal rigid rod. Besides gravitational effects, the expressions that we derived take proper acceleration of the resonator into account. As well as empty optical resonators, we considered optical resonators filled with a homogeneous dielectric material.

Our investigation revealed three fundamentally different effects. One is a simple gravitational red shift: the resonator is an extended object and time runs differently at different points inside the resonator. Therefore, the resonance frequencies of the resonator are not a global property of the resonator, but depend also on the position inside the resonator at which it is measured. The second effect is due to the difference between proper length and radar length, which leads to a shift of the resonance frequencies in the presence of non-zero curvature and acceleration even for a Born rigid resonator. The third effect is the deformation of the resonator due to curvature and acceleration, when the resonator is deformable. The deformation of the resonator is governed by only one parameter, the speed of sound $c_s$ in the rod. It turns out that the effects of deformations are larger than the relativistic effects, red shift and difference between proper length and radar length, by a factor $c^2/c_s^2$. A causal rigid rod can be considered to be one with the speed of sound equivalent to the speed of light, overcoming the problems of Born rigidity \cite{Natario:2014koa}. We gave an expression for the resonance frequency spectrum of a causal rigid rod in Eq. (\ref{eq:fundfreqdefocmcaus}). Since the largest speed of sound in any material is still many orders smaller than the speed of light, the deformations of realistic materials will dominate over the relativistic effects significantly. Therefore, a very high degree of control over the material parameters would be necessary to observe the relativistic effects.

Regardless of the suppression of relativistic effects on the resonator, the results derived in this article can be applied to general spacetime geometries if acceleration and tidal forces in the proper detector frame of the resonator are small enough. This includes freely falling resonators in strong gravitational fields like a black hole beyond the Schwarzschild radius or a uniformly accelerated cavity which we gave as examples in this article.
As a third example calculation, we considered the gravitational effect of an oscillating tungsten or gold sphere on the resonance frequencies of an optical resonator in Sec. \ref{sec:examplemass}. This situation is similar to the one considered in \cite{Schmoele:2017diss} and \cite{Schmoele:2016mde}, where the resonator is a second massive sphere on a support with a restoring force.

The precision of metrological experiments with resonators depends strongly on the knowledge of the resonance frequencies of these resonators. On the one hand, the effects of acceleration and curvature on the resonance frequencies can be seen as an experimental systematic error which has to be taken into account. On the other hand, these effects can be used to measure a proper acceleration or spacetime curvature. In such experimental situations, the model we used will certainly not be fully valid and the effects have to be calculated for the precise apparatus that is used. However, the results of this article can serve as a basis for investigations of the accessibility of spacetime parameters and parameters of states of motion in the more advanced framework of quantum metrology \cite{Howl:2016ryt}. 

In our analysis, the only non-Newtonian effects are the relativistic red shift and time dilation and the difference between radar length and proper length. However, the formalism employed here contains further relativistic effects (see table I of \cite{Ni1978proper}) such as the Sagnac effect and magnetic type gravitational effects such as frame dragging, which induces the Lens-Thirring effect in gyroscopes. It would be interesting to include these effects in a more detailed analysis. One way could be an extension to $3$-dimensional optical resonator geometries and the inclusion of the polarization of the light field. 

In the future, it would be desirable to have a description beyond the restrictions to small accelerations and curvatures. For that purpose, a fully relativistic description of elasticity has to be used such as those presented in \cite{Herglotz:1911mec}, \cite{Ehlers:1977dyn}, and \cite{Natario:2014koa}. For significant variations of the curvature on the length scale of the wave length of the resonator modes, it would be necessary to abandon the eikonal approximation and to derive the resonance frequencies directly from solutions of the Maxwell equations in a curved spacetime. This is the case if the effect of the gravitational field of the light inside the resonator is to be considered in full generality \cite{Braun:2015twa}. Furthermore, the effect of rotation of the resonator has to be considered in the future. This can be done by considering higher orders of the eikonal expansion or using methods of electrodynamics like the paraxial approximation.

\begin{acknowledgments}
We thank Ralf Menzel, Jonas Schm\"ole, Tobias Westphal, Philipp Haslinger, Jos\'e Nat\'ario, Luis Cort\'es Barbado, Jan Kohlrus, Ana Luc\'ia Baez and Uwe R. Fischer  for interesting remarks and discussions and Kiri Mochrie for writing assistance. D.R. thanks the Humboldt Foundation for funding his research in Vienna with their Feodor-Lynen Felloship. R.H. and I.F. would like to acknowledge that this project was made possible through the support of the grant ``Leaps in cosmology: gravitational wave detection with quantum systems'' (No. 58745) from the John Templeton Foundation. The opinions expressed in this publication are those of the authors and do not necessarily reflect the views of the John Templeton Foundation. I.F. would like to acknowledge that this project was made possible through the support of the grant ``Quantum Observers in a Relativistic World" from FQXi's Physics of the Observer program. The authors thank Jos\'e Ignacio Latorre and the Centro de Ciencias de Benasque Pedro Pascual for hosting the workshop ``Gravity in the Lab" in 2016 where the collaboration started that led to this article. The publication of this article is funded by the Open Access Publishing Fund of the University of Vienna.
\end{acknowledgments}

\appendix

\section{Relation to the concept of a rigid rod in special relativity}
\label{sec:specialrel}

In special relativity, the proper length of a rod is given as the coordinate distance between its endpoints, calculated in the coordinate system defined by the rest frame of the rod. Here, we call $L_p(s_\varrho)$ the proper length of the rod and describe it in the following. By definition, for every $\varrho_0$ and every point $s_{\varrho_0}(\varsigma_0)$ of the space-like curve $s_{\varrho_0}(\varsigma)$ representing the rod, there is a space-like tangent $s'_{\varrho_0}(\varsigma_0):=ds_{\varrho_0}(\varsigma)/d\varsigma|_{\varsigma_0}$. For every point of the curve $s_{\varrho_0}(\varsigma)$ representing the rod, there is an associated vector in the tangent space $T_{s_{\varrho_0}(\varsigma_0)}\mathcal{M}$ via the inverse of the exponential map, where the exponential map is given as $\mathrm{exp}_{s_{\varrho_0}(\varsigma_0)}: T_{s_{\varrho_0}(\varsigma_0)}\mathcal{M}\rightarrow \mathcal{M}$ and $\exp_{s_{\varrho_0}(\varsigma_0)}((\varsigma-\varsigma_0) s'_\varrho(\varsigma_0))=s_{\varrho_0}(\varsigma)$. In particular, the two endpoints of the rod $s_{\varrho_0}(a)$ and $s_{\varrho_0}(b)$ are associated with the vectors $(\varsigma_0-a)s'_{\varrho_0}(\varsigma_0)$ and $(b-\varsigma_0)s'_{\varrho_0}(\varsigma_0)$. Since $s_{\varrho_0}(\varsigma)$ is a space-like geodesic (in the sense of the auto-parallel property), the proper distance from $s_{\varrho_0}(\varsigma_0)$ to $s_{\varrho_0}(a)$ and $s_{\varrho_0}(b)$ is equivalent to the norm of $-(\varsigma_0-a)s'_{\varrho_0}(\varsigma_0)$ and $(b-\varsigma_0)s'_{\varrho_0}(\varsigma_0)$, respectively, with respect to the metric $g_{\mu\nu}$ at $s_{\varrho_0}(\varsigma_0)$. Hence, for every point $s_{\varrho_0}(\varsigma_0)$ on the rod, there is a representation of the rod as a straight line $\varsigma s'_{\varrho_0}(\varsigma_0)$ in the tangent space to this point and the sum of the proper distances in both directions of the rod is equivalent to the length of the line given as $(b-a)g_{s_{\varrho_0}(\varsigma_0)}(s'_{\varrho_0}(\varsigma_0),s'_{\varrho_0}(\varsigma_0))$. We can find coordinates such that $(g_{s_{\varrho_0}(\varsigma_0)})_{\mu\nu}=\eta_{\mu\nu}$. This is called a local Lorentz frame at $s_{\varrho_0}(\varsigma_0)$. In the local Lorentz frame, the coordinate distance (in tangent space) between the endpoints of the line $\varsigma s'_{\varrho_0}(\varsigma_0)$ is equivalent to its length $(b-a)g_{s_{\varrho_0}(\varsigma_0)}(s'_{\varrho_0}(\varsigma_0),s'_{\varrho_0}(\varsigma_0))$. In special relativity, the spacetime and the tangent space to every point can be identified since spacetime is flat. Then, the length of the line representing the rod in tangent space is also the proper length of the rod. Therefore, we can identify $L_p(s_\varrho)$ as the generalization of the proper length of a rigid rod in GR.

\section{Boundary conditions}
\label{sec:boundary}

In the following, we will will apply Maxwell's equations to the eikonal expansion in Eq. (\ref{eq:Amueikonal}) along the same lines as in \cite{Straumann:2012gen}. We will write $\nabla_\mu \zeta^\rho = {\zeta^\rho}_{;\mu}$ for the covariant derivative. In the following, we will apply the Lorenz gauge condition and Maxwell's equations to the eikonal expansion in Eq. (\ref{eq:Amueikonal}). Maxwell's equations can be written as \cite{Straumann:2012gen}
\begin{equation}
	{F_{\mu\nu;\lambda}}^{;\lambda} + ({R^\sigma}_\mu F_{\nu\sigma} - {R^\sigma}_\nu F_{\mu\sigma}) + R_{\alpha\beta\mu\nu} F^{\alpha\beta} = 0\,,
\end{equation}
where $R_{\mu\nu}$ is the Ricci tensor. We have
\begin{eqnarray}
	F_{\mu\nu;\lambda} &=& \mathrm{Re}\left(e^{i \frac{\alpha}{\lambda} S(x)}\sum_{n=0}^{\infty} \left(i\frac{\alpha}{\lambda}  \phi_{n,\mu\nu}\hat\xi_\lambda + \phi_{n,\mu\nu;\lambda}\right) \left(\frac{\lambda}{\alpha}\right)^n\right)\quad\rm{and}\\
	{F_{\mu\nu;\lambda}}^{;\lambda} &=& g^{\lambda\sigma} \mathrm{Re}\left(e^{i \frac{\alpha}{\lambda} S(x)}\sum_{n=0}^{\infty} \left(-\left(\frac{\alpha}{\lambda}\right)^2 \phi_{n,\mu\nu} \hat\xi_\lambda \hat\xi_\sigma + i\frac{\alpha}{\lambda} \left( \phi_{n,\mu\nu}\hat\xi_{\lambda;\sigma} + 2 \phi_{n,\mu\nu;\lambda}\hat\xi_{\sigma}\right) + \phi_{n,\mu\nu;\lambda\sigma}\right) \left(\frac{\lambda}{\alpha}\right)^n\right)\,.
\end{eqnarray}
In leading order, we find the null condition $g^{\lambda\sigma} \hat\xi_\lambda \hat\xi_\sigma=0$. By taking the covariant derivative of the null condition and taking into account that $\hat\xi_\mu = \partial_\mu S(x)$, we find 
\begin{equation}
	0=(g^{\lambda\sigma} \hat\xi_\lambda \hat\xi_\sigma)_{;\mu}=2\hat\xi^\sigma \hat\xi_{\sigma;\mu} = 2\hat\xi^\sigma S(x)_{;\sigma\mu} = 2\hat\xi^\sigma \hat\xi_{\mu;\sigma}
\end{equation}
which means that the integral curves of the vector field $\hat\xi^\sigma$ are light like geodesics. These are the light rays of geometrical optics. In the next to leading order, we find 
\begin{equation}
	0 = \phi_{0,\mu\nu}\hat{\xi}^\lambda_{\,\,\,;\lambda} + 2 \phi_{0,\mu\nu;\lambda}\hat\xi^{\lambda}
\end{equation}
We define the scalar
\begin{equation}
	\phi_0:=\left(g^{\alpha\gamma}g^{\beta\delta}\phi_{0,\alpha\beta}\phi^*_{0,\gamma\delta}\right)^{1/2}\,,
\end{equation}
and the polarization tensor $f_{0,\mu\nu}=\phi_{0,\mu\nu}/\phi_0$. We find that
\begin{eqnarray}
	\hat\xi^{\lambda} f_{0,\mu\nu;\lambda} &=& \hat\xi^{\lambda}\left((\phi_0)^{-1} \phi_{0,\mu\nu;\lambda} - (\phi_0)^{-2}\phi_{0,\mu\nu}\phi_{0;\lambda}\right)\\
	&=& \hat\xi^{\lambda}\left((\phi_0)^{-1} \phi_{0,\mu\nu;\lambda} - \frac{1}{2}(\phi_0)^{-3}\phi_{0,\mu\nu}g^{\alpha\gamma}g^{\beta\delta}\left(\phi_{0,\alpha\beta;\lambda}\phi^*_{0,\gamma\delta}+\phi^*_{0,\alpha\beta;\lambda}\phi_{0,\gamma\delta}\right)\right)\\
	&=& (\phi_0)^{-1} \phi_{0,\mu\nu;\lambda}\hat\xi^{\lambda} + \frac{1}{2}(\phi_0)^{-1}\phi_{0,\mu\nu}\hat{\xi}^\lambda_{\,\,\,;\lambda}=0\,.
\end{eqnarray}
This means that the zeroth order polarization tensor is parallel transported along the light rays. Furthermore, for linear polarization, we can write $f_{0,\mu\nu}=\exp(i\varphi_0)\bar f_{0,\mu\nu}$, where $\phi$ and $\bar f_{0,\mu\nu}$ are real. From $ f_{\mu\nu;\lambda} \hat\xi^\lambda =  0$, we find that $\varphi_{0,\lambda} \hat\xi^\lambda =0$. Therefore, the phase of the zeroth order amplitude function does not change along the light ray. In particular, we can assume that $\phi_{0,\mu\nu}$ is real everywhere as we can set the initial conditions accordingly.


With these considerations, we can investigate the boundary conditions at the mirrors. To express the boundary conditions in a covariant form, we define the frames of the mirrors in the following. The tangents $\dot\gamma_\rm{A}(\varrho)^\mu$ and $\dot\gamma_\rm{B}(\varrho)^\mu$ of the world lines of the mirrors define a spacetime split; the spatial slice at the mirror $(i)=\rm{A,B}$ is defined as the set of vectors $r^{(i)\mu}$ such that $g_{\mu\nu}r^{(i)\mu}\dot\gamma_{(i)}(\varrho)^\nu=0$ (no summation of $i$). Inside these spatial slices, we can define three orthonormal vectors $\epsilon^{(i)\mu}_j$ such that the vector $ \epsilon^{(i)\mu}_3$ is orthogonal to the mirror and the normal vectors $ \epsilon_{1}^{(i)\mu}$ and $ \epsilon_{2}^{(i)\mu}$ are tangential to the mirror\footnote{We only need the latter to be defined up to rotations around $ \epsilon^{(i)\mu}_1$ in the spatial slice.}. Furthermore, we choose $ \epsilon_{1}^{(i)\mu}$ to be directed in the polarization direction of the right propagating light field at the mirror $(i)$. Together with $\epsilon^{(i)\mu}_0 = \dot\gamma_{(i)}(\varrho)^\mu/|\dot\gamma_{(i)}(\varrho)|$, the vectors $\epsilon^{(i)\mu}_J$ ($J\in\{1,2,3\}$) form an orthonormal tetrad. 
Using the tetrads, the components of the field strength tensor in the frame of the mirror are given as $F^{(i)}_{\mathcal{M}\mathcal{N}}(\varrho)=\epsilon_{\mathcal{M}}^{(i)\mu}\epsilon_{\mathcal{N}}^{(i)\nu}F_{\mu\nu}(\gamma_{(i)}(\varrho))$. Then, the boundary conditions at the mirrors are that the electric field is perpendicular and the magnetic field parallel to the mirrors, i.e. $F^{(i)}_{01}(\varrho)= 0 = F^{(i)}_{02}(\varrho)$ and $F^{(i)}_{12}(\varrho)=0$. 

The tetrads were defined such that the polarization direction of the light field is in the direction of $ \epsilon_{1}^{(i)\mu}$. We define $\phi^{(i)r/l}_{n,01}(\varrho):=\epsilon_{0}^{(i)\mu}\epsilon_{1}^{(i)\nu}\phi^{r/l}_{n,\mu\nu}(\gamma_{(i)}(\varrho))$ which are non-zero and we find the boundary conditions
\begin{eqnarray}
	0=F_{01}^{(i)\mathrm{res}}(\varrho)=\mathrm{Re}\left( e^{i \frac{\alpha}{\lambda} S^r(\gamma_{(i)}(\varrho))}\sum_{n=0}^{\infty} \phi^{(i)r}_{n,01}(\varrho) \left(\frac{\lambda}{\alpha}\right)^n + e^{i \frac{\alpha}{\lambda} S^l(\gamma_{(i)}(\varrho))}\sum_{n=0}^{\infty} \phi^{(i)l}_{n,01}(\varrho) \left(\frac{\lambda}{\alpha}\right)^n\right)\,.
\end{eqnarray}
From the lowest order in $\lambda/\alpha$, we find that 
\begin{eqnarray}\label{eq:boundcond}
	0=\mathrm{Re}\left( e^{i \frac{\alpha}{\lambda} S^r(\gamma_{(i)}(\varrho))} \phi^{(i)r}_{0,01}(\varrho) + e^{i \frac{\alpha}{\lambda} S^l(\gamma_{(i)}(\varrho))}\phi^{(i)l}_{0,01}(\varrho)\right)\,.
\end{eqnarray}
Above, we found that the zeroth order amplitude tensors are real. Then, the boundary condition (\ref{eq:boundcond}) can only be fulfilled for all $\varrho$ if $\phi^r_{0,01}(\varrho)=\phi^l_{0,01}(\varrho)$ and $\frac{\alpha}{\lambda} S^r(\gamma_{(i)}(\varrho))=\frac{\alpha}{\lambda} S^l(\gamma_{(i)}(\varrho)) + 2\pi m_{(i)}$, where $m_{(i)} \in \mathbb{Z}$.

\section{Deformations of a rod}
\label{sec:deformations}

For isotropic media, the stiffness tensor depends only on the Young's modulus $Y$, the shear modulus $G$ and the Poisson ratio $\nu$. We have
\begin{eqnarray}
	\varepsilon_{xx} &=& \frac{1}{Y}\left(\sigma_{xx} - \nu (\sigma_{yy}+\sigma_{zz})\right)\\
	\varepsilon_{yy} &=& \frac{1}{Y}\left(\sigma_{yy} - \nu (\sigma_{xx}+\sigma_{zz})\right)\\
	\varepsilon_{zz} &=& \frac{1}{Y}\left(\sigma_{zz} - \nu (\sigma_{xx}+\sigma_{yy})\right)\\
	\varepsilon_{ij} &=& \varepsilon_{ji} = \frac{1}{2G}\sigma_{ij}\,\quad\rm{for}\,\,i\neq j\,.
\end{eqnarray}
Since the change of thickness of the rod holding the resonator and its deformations in the $x$-$y$-plane are not of interest for us, we can restrict our considerations to $\varepsilon_{zz}$, $\varepsilon_{xz}$ and $\varepsilon_{yz}$. The elements of the strain tensor $\varepsilon_{xz}$ and $\varepsilon_{yz}$ lead to a deformation of the curve $s(\varsigma)$ in the $x$ and $y$ direction, respectively. Since the corresponding forces are always transversal to the line elements of the rod, they only bend the rod and do not change its proper length.  In the proper detector frame, the proper length of the part of the rod in the positive $z$-direction of the support is approximately given as
\begin{eqnarray}\label{eq:triangledef}
	\nonumber \frac{1+\beta}{2}L_p &\approx & \int d\varsigma\, ((s^{\prime\, x})^2 + (s^{\prime\, y})^2 + (s^{\prime\, z})^2)^{1/2} \\
		&\approx & \int_0^{(1+\beta)L_p/2 - \delta b} dz\left( 1 + \frac{1}{2}\left(\left(\frac{s^{\prime\, x}}{ s^{\prime\, z}}\right)^2 + \left(\frac{s^{\prime\, y}}{ s^{\prime\, z}}\right)^2\right)\right)\,,
\end{eqnarray}
where $\delta b$ is the shift of the $z$-coordinate of the position of mirror B. For the analysis of the transversal deformations, let us assume that the rod has a rectangular cross section with side lengths $w_x$ and $w_y$. Furthermore, let us consider the extreme case of $\beta = 1$. An expression for the transversal deformation of such a rod can be found, for example, in Eq. 2.2  \cite{Srivastava:2006the}. For the $x$-direction, we find 
\begin{equation}
	\frac{d^2 s^x}{d z^2} \le 6 \frac{\rho}{Y}\mathbf{a}^{x}_{P\rm{max}} \frac{(L_p-z)^2}{w_x^2}
\end{equation}
where $\mathbf{a}^{x}_{P\rm{max}}$ is the maximal acceleration in $x$-direction experienced by
a part of the rod. With $s' = ds/d\varsigma = 0$ at $z=0$, we obtain that 
\begin{equation}\label{eq:dxoverdzdef}
	\frac{s^{\prime\,x}}{s^{\prime\,z}} = \frac{d s^x}{d z} \le 2 \frac{\rho}{Y}\mathbf{a}^{x}_{P\rm{max}} \frac{L_p^3 - (L_p-z)^3}{w_x^2}\,.
\end{equation}
A similar expression can be found for $s^{\prime\,y}/s^{\prime\,z}$. With Eq. (\ref{eq:triangledef}), we obtain the approximate upper bounds for the change of the $z$-position of the mirror B
\begin{equation}\label{eq:deltaLptrans}
	\delta b \le \frac{9}{7}\frac{L_p^7}{c_s^4} \left(\left(\frac{\mathbf{a}^{x}_{P\rm{max}}}{w_x^2}\right)^2 + \left(\frac{\mathbf{a}^{y}_{P\rm{max}}}{w_y^2}\right)^2\right)\,.
\end{equation}
Then, the new position of mirror B is approximately $(s^x(L_p), s^y(L_p), L_p - \delta b)$, where we get
\begin{eqnarray}
	s^x(L_p) \leq \frac{3L_p^4}{2c_s^2} \frac{\mathbf{a}^{x}_{P\rm{max}}}{w_x^2} \quad \rm{and}\quad s^y(L_p) \leq \frac{3L_p^4}{2c_s^2} \frac{\mathbf{a}^{y}_{P\rm{max}}}{w_y^2}\,,
\end{eqnarray}
by integration Eq. (\ref{eq:dxoverdzdef}) and the corresponding expression for the $y$-direction. Since $\delta b$, $s^x(L_p)$ and $s^y(L_p)$ are already of second and first order in the metric perturbation, respectively, the change of the round trip time can be calculated as
\begin{eqnarray}
	\delta T &\approx & \frac{1}{c}\left(\left((L_p-\delta b)^2 + s^x(L_p)^2 + s^y(L_p)^2\right)^{1/2}-L_p\right)\\
	&\approx & - \frac{1}{3c}\frac{L_p^7}{c_s^4} \left(\left(\frac{\mathbf{a}^{x}_{P\rm{max}}}{w_x^2}\right)^2 + \left(\frac{\mathbf{a}^{y}_{P\rm{max}}}{w_y^2}\right)^2\right)
\end{eqnarray}
Let us define $\mathbf{a}_{P,\rm{av}}^z$ as the larger of the values of $\langle\beta|\mathbf{a}^z(\tau)|\rangle$ and $\langle(3\beta^2+1)L_pc^2|R_{0z0z}(\tau)|/6\rangle$, where $\langle \rangle$ denotes the averaging over the interaction time. Comparison of Eq. (\ref{eq:deltaLptrans}) with Eq. (\ref{eq:length}) shows that the effect of the transversal bending on the length of the rod can be neglected in comparison to the effect of the longitudinal deformations if 
\begin{eqnarray}\label{eq:condtrans}
	\mathbf{a}_{P,\rm{av}}^z &\gg &  \rm{max}\left\{ L_p^5\langle |\mathbf{a}^{x}_{P\rm{max}}|\rangle^2/c_s^2 w_x^4 , L_p^5\langle |\mathbf{a}^{y}_{P\rm{max}}|\rangle^2/c_s^2 w_y^4 \right\}\,.
\end{eqnarray}
In the gravitational field of a small massive sphere of $100\,\rm{g}$ of the example in Sec. \ref{sec:examplemass}, an observer at rest experiences an acceleration of the order of $10^{-10}\,\rm{ms^{-2}}$. So we assume $\mathbf{a}_{P,\rm{av}}^z = 10^{-10}\,\rm{ms^{-2}}$, $\langle |\mathbf{a}^{x}_{P\rm{max}}|\rangle\le 10^{-10}\,\rm{ms^{-2}}$ and $\langle |\mathbf{a}^{y}_{P\rm{max}}|\rangle\le 10^{-10}\,\rm{ms^{-2}}$ \footnote{We consider the massive sphere as the only source of a gravitational field here. In an earthbound laboratory, the effect of the Earth's gravitational field has to be taken into account as well.}. Let us consider an aluminum rod where $c_s= 5\times 10^{3}\rm{ms^{-1}}$. For a rod of length $1\,\rm{cm}$, we find that $L_p/w_x\le 10^3$ and $L_p/w_y\le 10^3$ is sufficient to fulfill the conditions in Eq. (\ref{eq:condtrans}). Let us consider the situation for accelerations of the order of $10\,\rm{ms^{-2}}$ as they are experienced in the gravitational field of the earth. So we assume $\mathbf{a}_{P,\rm{av}}^z = 10\,\rm{ms^{-2}}$, $\langle |\mathbf{a}^{x}_{P\rm{max}}|\rangle\le 10\,\rm{ms^{-2}}$ and $\langle |\mathbf{a}^{y}_{P\rm{max}}|\rangle\le 10\,\rm{ms^{-2}}$. For an aluminum rod of length $10\,\rm{cm}$, the conditions in Eq. (\ref{eq:condtrans}) are fulfilled for $L_p/w_x\le 10$ and $L_p/w_y\le 10$. For larger accelerations, the orientation has to be chosen such that $\mathbf{a}^{x}_{P\rm{max}}\ll \mathbf{a}_{P,\rm{av}}^z$ and $\mathbf{a}^{y}_{P\rm{max}}\ll \mathbf{a}_{P,\rm{av}}^z$ to fulfill the conditions and still use a rod.

Now, let us consider the longitudinal deformation. From $\mathbf{a}_P^j\approx -c^2\Gamma_{00}^j\approx c^2\partial_j h_{00}$, we obtain the inertial and tidal forces on the rod by multiplication with the mass density $\rho$. Since $h_{00}$ contains terms that are independent of $z$ and terms that are proportional to $z$ and $z^2$, we can write the acceleration as 
\begin{equation}
	\mathbf{a}_P^z(\tau,x,y,z) = \mathbf{a}_{P}^z(\tau,x,y,0) + z \left.\frac{d}{dz}\mathbf{a}_{P}^{z}(\tau,x,y,z)\right|_{z=0}\,.
\end{equation}
Let us assume that the rod has a constant cross section $A$ and a constant mass density. Then, the sum of inertial forces and gravitational force along the rod acting on a segment of the rod at $z>0$ can be approximated as
\begin{eqnarray}
	\nonumber F_+^{z}(\tau,z)
	&\approx &	\int_z^b dz' A\rho \,\mathbf{a}_P^z(\tau,0,0,z')\\
	&\approx & (b-z) A\rho\,\mathbf{a}_{P}^z(\tau,0,0,0) + \frac{1}{2}(b^2-z^2)  A\rho\left.\frac{d}{dz}\mathbf{a}_{P}^{z}(\tau,0,0,z)\right|_{z=0}\,.
\end{eqnarray}
where, by considering the acceleration only at $x=0=y$, we neglected terms proportional to the width of the rod. For the force along the rod acting on a segment of the rod at $z<0$, we find
\begin{eqnarray}
	F_-^{z}(\tau,z) \approx  (z-a) A\rho\,\mathbf{a}_{P}^z(\tau,0,0,0) + \frac{1}{2}(z^2-a^2)  A\rho\left.\frac{d}{dz}\mathbf{a}_{P}^{z}(\tau,0,0,z)\right|_{z=0}\,.
\end{eqnarray}
Due to the support, this corresponds to the stresses
\begin{equation}
	\sigma^\pm_{zz}(\tau,z) = \pm \frac{F^z_\pm(\tau,z)}{A}\,.
\end{equation}

The differential force in the $x$-direction acting on a 1-dimensional segment of the rod with coordinates $x$, $y$ and $z$ induced by all 1-dimensional segments with the same $z$-coordinate, the same $y$-coordinate and $x'>x$ can be written as
\begin{eqnarray}
	\nonumber dF_+^{x}(\tau,x,y,z)
	& = & dz \int_x^{w_x/2} dx' w_y\rho \mathbf{a}_P^x(\tau,x',y,z) \,.
\end{eqnarray}
Furthermore, we find
\begin{eqnarray}
	\nonumber dF_-^{x}(\tau,x,y,z)
	& = & dz \int_{-w_x/2}^x dx' w_y\rho \mathbf{a}_P^x(\tau,x',y,z) \,.
\end{eqnarray}
for the differential force induced by all 1-dimensional segments with the same $z$-coordinate, the same $y$-coordinate and $x'<x$. Since the metric (\ref{eq:metricproper}) contains constant, linear and quadratic terms in the spatial coordinate and $\mathbf{a}_P^j\approx -c^2\Gamma_{00}^j$, we conclude that $d\mathbf{a}_{P}^{x}(\tau,x,y,z)/dx$ cannot depend on $y$ in first order in the metric perturbation, and we find that the acceleration in the $x$-direction can be written as
\begin{equation}\label{eq:accsep}
	\mathbf{a}_P^x(\tau,x,y,z) = \mathbf{a}_{P}^x(\tau,0,y,z) + x \frac{d}{dx}\left.\mathbf{a}_{P}^{x}(\tau,x,0,z)\right|_{x=0}\,.
\end{equation}
The first term corresponds to an acceleration that all segments feel in the same way. Therefore, it does not lead to a stress. Hence, the stress on a segment of the rod at $z$ becomes
\begin{eqnarray}
	\nonumber \sigma_{xx}(\tau,z)
	&=&	\frac{w_x^2}{8}\rho \left.\frac{d}{dx}\mathbf{a}_{P}^{x}(\tau,x,0,z)\right|_{x=0}\,.
\end{eqnarray}
An equivalent expression can be derived for the stress $\sigma_{yy}$. The length change of the rod is given as
\begin{equation}\label{eq:lengthtotal}
	\begin{array}{lcl}	
	\delta L_p(\tau)
	&\simeq&   \int_0^b dz' \varepsilon^+_{zz}(\tau,z') + \int^0_a dz' \varepsilon^-_{zz}(\tau,z')\\
	&=& \frac{1}{Y}\int_0^b dz' \sigma^+_{zz}(\tau,z') + \frac{1}{Y}\int^0_a dz' \sigma^-_{zz}(\tau,z')	-\frac{\nu}{Y}\int_a^b dz'\,\left(\sigma_{xx}(\tau,z') + \sigma_{yy}(\tau,z')\right)
	\end{array}
\end{equation}
We obtain that 
\begin{eqnarray}
	&&\frac{1}{Y}\int_0^b dz' \sigma^+_{zz}(\tau,z') + \frac{1}{Y}\int^0_a dz' \sigma^-_{zz}(\tau,z')\\
	&=&\frac{\rho}{Y}\left( \frac{1}{2}(b^2-a^2) \mathbf{a}_{P}^z(\tau,0,0,0) + \frac{1}{3}(b^3-a^3) \left.\frac{d}{dz}\mathbf{a}_{P}^{z}(\tau,0,0,z)\right|_{z=0}\right)\\
	&=&\frac{\rho}{Y}L_p\left( \frac{1}{2}\beta L_p \mathbf{a}_{P}^z(\tau,0,0,0) + \frac{1}{12}(3\beta^2+1)L_p^2 \left.\frac{d}{dz}\mathbf{a}_{P}^{z}(\tau,0,0,z)\right|_{z=0}\right)\,.
\end{eqnarray}
Since the highest polynomial order of terms in the metric perturbation in the coordinates is 2, $\frac{d}{dx}\mathbf{a}_{P}^{x}(\tau,x,0,z)|_{x=0}$ can only contain terms that are independent of $z$ and terms that are linear in $z$. Hence, we find 
\begin{eqnarray}
	&&\frac{\nu}{Y}\int_a^b dz'\,\left(\sigma_{xx}(\tau,z') + \sigma_{yy}(\tau,z')\right)\\
	&\approx& \frac{\nu\rho}{Y}L_p\left(\frac{w_x^2}{8}\left.\frac{d}{dx}\mathbf{a}_{P}^{x}(\tau,x,0,0,0)\right|_{x=0} + \frac{w_y^2}{8}\left.\frac{d}{dy}\mathbf{a}_{P}^{y}(\tau,0,y,0)\right|_{y=0} \right)\,.
\end{eqnarray}
Therefore, the effect of acceleration and curvature on the proper length via $\sigma_{xx}$ and $\sigma_{yy}$ is suppressed by a factor $\nu w_x/L_p$ and $\nu w_y/L_p$, respectively, in comparison to the effect via $\sigma_{zz}$. For most materials $\nu < 1$ and we can assume that $w_x/L_p\ll 1$. Therefore, if $\beta\mathbf{a}_{P}^z(\tau,0,0,0)$ or $(3\beta^2+1)L_p \frac{d}{dz}\mathbf{a}_{P}^{z}(\tau,0,0,z)|_{z=0}/6$ is of the same order or larger than $w_x\frac{d}{dx}\mathbf{a}_{P}^{x}(\tau,x,0,0)_{x=0}/4$ and $w_y\frac{d}{dy}\mathbf{a}_{P}^{y}(\tau,0,y,0)|_{y=0}/4$ and if the oscillations of the transversal stresses are not on resonant with any elastic mode of the rod that the longitudinal stresses are not on resonance with, we can neglect the effect of the transversal stresses and we can restrict our considerations to $\sigma^+_{zz}$ and $\sigma^-_{zz}$. Then, we can write the conditions as
\begin{eqnarray}
	\mathbf{a}_{P,\rm{av}}^z &\ge &  \rm{max}\left\{ w_x c^2 \langle |R_{0x0x}|\rangle, w_y c^2 \langle |R_{0y0y}|\rangle\right\}\,,
\end{eqnarray}

\section{The causal deformable rod from relativistic elasticity}
\label{sec:causal}

In \cite{Natario:2014koa}, a covariant formulation of the relativistic elastic rod was given. In this section, we show that the definitions of \cite{Natario:2014koa} lead to our result Eq. (\ref{eq:fundfreqdefocmcaus}) for the causal rigid rod when applied to the metric in Eq. (\ref{eq:metricproper}) in the proper detector frame. 

The author of \cite{Natario:2014koa} formulates the theory of 1-dimensional relativistic elastic bodies by considering a motion of a 1-dimensional continuum moving in a $1+1$-dimensional spacetime.   Our arguments from Sec. \ref{sec:rigidrod}, \ref{sec:fundfreqrigid} and \ref{sec:realrod} lead exactly to such a situation. The rod is dragged along the world line of its support or its center of mass is assumed to move along a geodesic. All accelerations of the rod segments are encoded in the metric in the proper detector frame given by Eq. (\ref{eq:metricproper}). Furthermore, our rod is assumed to lie along a spatial geodesic and we neglect all transversal accelerations. What remains is only gravitational effects along the rod encoded by the metric corresponding to the line element
\begin{equation}
	ds^2 = -(1-h^\rm{P}_{00}(\tau,z)) d\tau^2 + dz^2\,.
\end{equation}
Due to our assumption that acceleration and curvature only change very slowly, we find that this situation corresponds to Eq. (22) of \cite{Natario:2014koa}. The coordinate transformation in Eq. (23) of \cite{Natario:2014koa}, $\tilde{z}=f(z)$ with $f(z)=\int_0^z dz'\,(1-h^\rm{P}_{00})^{-1/2}$  leads to 
\begin{equation}
	ds^2 \approx -(1-h^\rm{P}_{00}(\tau,f^{-1}(\tilde{z}))) \left(d\tau^2 + d\tilde{z}^2\right) \approx (1-h^\rm{P}_{00}(\tau,\tilde{z})) \left(-d\tau^2 + d\tilde{z}^2\right)  \,.
\end{equation}
in first order in the metric perturbation since $f^{-1}(\tilde{z}) = \tilde{z}$ in zeroth order in the metric perturbation. The rigid rod of \cite{Natario:2014koa} has constant coordinate length in the coordinates $(\tau,\tilde{z})$, which are called conformal coordinates because the line element differs from the that of Minkowski space only by a conformal factor $e^{2\phi(\tilde{z})}$, where in our case, $e^{2\phi(\tilde{z})}=(1-h^\rm{P}_{00}(\tau,\tilde{z}))$. This rigid rod can be called a causal rigid rod because the speed of sound in the rod material is equivalent to the speed of light. In contrast, a Born rigid rod would correspond to an infinite speed of sound. 

The square root of the conformal factor is the stretch constant of \cite{Natario:2014koa}. We obtain the proper length of the causal rigid rod by integrating the stretch constant from one end of the rod to the other. However, we have to note that the stretch factor also contains boundary conditions of the rod; every point at which $\phi(\tilde{z})$ vanishes corresponds to a free end of the rod. Therefore, we cannot just use the expression for $h^\rm{P}_{00}$ that we used in Sec. \ref{sec:realrod}. We have to consider the two sides of our rod separately, and in each situation, add a constant to $h^\rm{P}_{00}$ such that the free end is at $a$ or $b$. Adding a constant to the metric does not change any dynamics and we are free to do such an operation. We define
\begin{eqnarray}
	h^\rm{A}_{00}(\tau,\tilde{z})&:=&h^\rm{P}_{00}(\tau,\tilde{z})-h^\rm{P}_{00}(\tau,a)\,\quad\rm{and}\\
	h^\rm{B}_{00}(\tau,\tilde{z})&:=&h^\rm{P}_{00}(\tau,\tilde{z})-h^\rm{P}_{00}(\tau,b)\,.
\end{eqnarray}
The proper length becomes
\begin{eqnarray}
	\nonumber L_p &=& \int_{a}^0 d\tilde{z}\,(1-h^\rm{A}_{00}(\tau,\tilde{z}))^{1/2} +  \int_{0}^b dy\,(1-h^\rm{B}_{00}(\tau,\tilde{z}))^{1/2}\\
		&\approx & \int_{a}^0 d\tilde{z}\,\left(1 + \frac{\mathbf{a}^z}{c^2}(\tilde{z}-a) + \frac{R_{0z0z}}{2}(\tilde{z}^2-a^2)  \right) +  \int_{0}^b d\tilde{z}\,\left(1 + \frac{\mathbf{a}^z}{c^2}(\tilde{z}-b) + \frac{R_{0z0z}}{2}(\tilde{z}^2-b^2)  \right)
\end{eqnarray}
and we reproduce the result of Eq. (\ref{eq:length}) for $c_s = c$.

\bibliography{frequency_in_gr}

\end{document}